* Manuscript

# Chaos synchronization of the master-slave generalized Lorenz systems via linear state error feedback control

Xiaofeng Wu*

*Center for Control and Optimization, South China University of Technology, Guangzhou 510640, P. R. China*
*and*
*Guangzhou Navy Academy, Guangzhou 510430, P. R. China*

Guanrong Chen

*Department of Electronic Engineering, City University of Hong Kong, Kowloon, Hong Kong, P. R. China*

Jianping Cai

*Department of Applied Mechanics and Engineering, Zhongshan University, Guangzhou 510275, P. R. China*

**Abstract**

This paper provides a unified method for analyzing chaos synchronization of the generalized Lorenz systems. The considered synchronization scheme consists of identical master and slave generalized Lorenz systems coupled by linear state error variables. A sufficient synchronization criterion for a general linear state error feedback controller is rigorously proven by means of linearization and Lyapunov's direct methods. When a simple linear controller is used in the scheme, some easily implemented algebraic synchronization conditions are derived based on the upper and lower bounds of the master chaotic system. These criteria are further optimized to improve their sharpness. The optimized criteria are then applied to four typical generalized Lorenz systems, *i.e.* the classical Lorenz system, Chen system, Lü system and a unified chaotic system, obtaining precise corresponding synchronization conditions. The advantages of the new criteria are revealed by analytically and numerically comparing their sharpness with that of the known criteria existing in the literature.

* Corresponding author: Xiaofeng Wu

Postal address: Guangzhou Navy Academy 075

Shisha Lu, Shijing, Baiyun Qu

Guangzhou 510430

P. R. China

Email address: wuxiaof@21cn.com, mathwxf@sina.com.

Tel: 86-20-86407656.

Fax: 86-20-61695540.

## 1. Introduction

The generalized Lorenz system is a class of three-dimensional (3-D) autonomous quadratic chaotic systems, first introduced in [1] and then extended and investigated in detail in [2] and [3]. Several well-known chaotic systems, e.g. the classical Lorenz system, Chen system, Lü system and a unified chaotic system [4], are special cases of the generalized Lorenz system. The introduction of the generalized Lorenz system makes possible the study of characteristics and dynamics of such typical chaotic systems in a unified manner.

As a topic of dynamical systems theory, chaos synchronization of generalized Lorenz systems is interesting and important due to its potentials in both theory and applications [2] [5]. On the other hand, linear state error feedback control is a robust and easily implemented control technique available for chaos synchronization. Recently, this technique has been widely used to synchronize two identical classical Lorenz systems [6]-[10], Chen systems [11], Lü systems [12] [13] and unified chaotic systems [7] [14]. Some sufficient synchronization conditions have been derived by means of the Lyapunov stability theory [6]-[14], however, generally providing rather conservative design for the linear controllers, as can be seen from [6] and Section 4 below.

This paper systematically investigates a synchronization scheme consisting of two identical generalized Lorenz systems unidirectionally coupled by a linear state error feedback controller. A sufficient synchronization criterion is first proven by means of linearization and Lyapunov's direct methods, and then applied to derive the algebraic synchronization conditions for some special (simple) linear controllers, where a useful piecewise function is introduced and the upper and lower bounds of the master chaotic system are utilized. These criteria are further optimized to achieve less conservative or sharper results. The new synchronization criteria for the classical Lorenz system, Chen system, Lü system and a unified chaotic system are finally developed based on the optimized results, and verified to be sharper than the existing ones proposed in [6]-[14] via both theoretical analysis and numerical simulation.

The main contributions of the paper can be summarized into two aspects. First, some sufficient synchronization criteria for the linearly coupled generalized Lorenz systems are obtained in an explicit algebraic form. Second, the new synchronization criteria for some typical chaotic systems are developed, generalizing and improving most, if not all, existing criteria of this type.

The rest of the paper is organized as follows. In the next section, a master-slave synchronization scheme for the generalized Lorenz systems coupled by a linear state error feedback controller is introduced. Section 3 presents the main theoretical results on the synchronization criteria for the scheme, including their proofs and optimization. In section 4, some algebraic synchronization criteria for several typical chaotic systems are derived and compared analytically and numerically with the existing criteria. Concluding remarks are finally given in the last section.

## 2. The Synchronization Scheme

Consider a class of generalized Lorenz systems described by [2]

$$\dot{y}=Ay+f(y)=\begin{pmatrix} a_{11} & a_{12} & 0 \\ a_{21} & a_{22} & 0 \\ 0 & 0 & a_{33} \end{pmatrix} y + y_1 \begin{pmatrix} 0 & 0 & 0 \\ 0 & 0 & -1 \\ 0 & 1 & 0 \end{pmatrix} y, \tag{1}$$

where $y=(y_1, y_2, y_3)^T$ and $A$ is a constant matrix of parameters.

According to the classification given in [2], the Lorenz system satisfies the condition $a_{12}a_{21}>0$, Chen system satisfies $a_{12}a_{21}<0$, and Lü system satisfies $a_{12}a_{21}=0$.

Four typical chaotic systems can be specified, as follows:

(i) the classical Lorenz system

$$a_{12}=-a_{11}=a,\ a_{21}=c,\ a_{22}=-1, \text{ and } a_{33}=-b; \tag{2}$$

(ii) the Chen system

$$a_{12}=-a_{11}=a,\ a_{21}=c-a,\ a_{22}=c, \text{ and } a_{33}=-b; \tag{3}$$

(iii) the Lü system

$$a_{12}=-a_{11}=a,\ a_{21}=0,\ a_{22}=c, \text{ and } a_{33}=-b; \tag{4}$$

(iv) a unified chaotic system

$$a_{12}=-a_{11}=25+\alpha,\ a_{21}=28-35\alpha,\ a_{22}=29\alpha-1 \text{ and } a_{33}=-\frac{8+\alpha}{3}, \tag{5}$$

where $a>0, b>0, c>0$ and $\alpha\in[0,1]$.

Now, construct a master-slave synchronization scheme for two identical generalized Lorenz systems coupled by a linear state error feedback controller $u(t)$, as follows:

$$\begin{cases} \text{Master: } \dot{x}=Ax+f(x), \\ \text{Slave: } \quad \dot{z}=Az+f(z)+u(t), \\ \text{Control: } u(t)=K(x-z), \end{cases} \tag{6}$$

where $x, z\in R^3$, the matrix $A$ and the nonlinearity $f(\cdot)$ are defined as in (1), and $K\in R^{3\times3}$ is a constant matrix, referred to as the "feedback gain matrix" or "coupling matrix".

Define the error variable $e=x-z$, where $e=(e_1,e_2,e_3)^T=(x_1-z_1,x_2-z_2,x_3-z_3)^T$. From (6), one can obtain a dynamical error system:

$$\dot{e}=(A-K)e+f(x)-f(z)=(A-K)e+f(x)-f(x-e). \tag{7}$$

The objective of the synchronization scheme (6) is to design a constant coupling matrix $K$ such that the trajectories, $x(t)$ and $z(t)$, of the master and slave systems satisfy

$$\lim_{t\to\infty}\|x(t)-z(t)\|=0, \tag{8}$$

where $\|\cdot\|$ denotes the Euclidean norm.

From the viewpoint of control theory, the synchronization issue (8) is equivalent to the uniform asymptotical stability of the error system (7) at $e=0$. Hence, chaos synchronization in the sense of (8) is referred to as *uniform chaos synchronization* below.

In order to find a sufficient synchronization criterion for the scheme (6), the following assumption on the master system is needed. This assumption is in light of the master system being free and chaotic, and based on a well-known fact that chaotic attractors are bounded in the phase space.

**Assumption.** In the synchronization scheme (6), the chaotic trajectory of the master system is bounded, *i.e.*, for any bounded initial state $x_0$ within the defining domain of the master system, there exist some finite real constants $\underline{\rho}_i(x_0)$ and $\overline{\rho}_i(x_0)$ such that

$$\underline{\rho}_i\le x_i(t,x_0)\le\overline{\rho}_i,\ i=1,2,3,\ \forall t\ge 0.$$

## 3. Main Theoretical Results

In this section, the generalized Lorenz system (1) is considered and some sufficient synchronization criteria for scheme (6) are proven and then optimized. These criteria can be used to analytically design various structures of the coupling matrix $K$ guaranteeing the scheme (6) to achieve uniform chaos synchronization.

For a $C^1$ nonlinear function $f: R^n \to R^m$, let

$$\frac{\partial}{\partial y_i} f_i(x) = \left.\frac{\partial f_i(y)}{\partial y_i}\right|_{y=x}, \quad i = 1, 2, \cdots, m, \quad j = 1, 2, \cdots, n.$$

Then, the Jacobian matrix of $f(y)$ given by (1) at the orbit $x$ of the master system equals

$$\frac{\partial}{\partial y} f(x) = \left(\frac{\partial}{\partial y_i} f_i(x)\right)_{3\times 3} = \begin{pmatrix} 0 & 0 & 0 \\ -x_3 & 0 & -x_1 \\ x_2 & x_1 & 0 \end{pmatrix}. \tag{9}$$

The following result provides a sufficient condition for designing the general coupling matrix $K$.

**Theorem 1.** The master-slave synchronization scheme (6) achieves uniform chaos synchronization if there exists a symmetric positive definite matrix $P \in R^{3\times 3}$ and a coupling matrix $K \in R^{3\times 3}$ such that

$$Y = \left(A - K + \frac{\partial}{\partial y} f(x)\right)^T P + P\left(A - K + \frac{\partial}{\partial y} f(x)\right) \tag{10}$$

is negative definite for all orbits $x \in R^3$ of the master system.

*Proof.* The relation between the stability of the error system (7) and that of its linearized error system is first discussed.

Since the nonlinear function $f_i(y)\,(i = 1,2,3)$ defined by (1) have continuous first and second partial derivatives for $y \in R^3$, the Taylor expansion of $f(y)$ at the orbits $x$ and $z$ of the master and slave system, respectively, is

$$f(z) = f(x) - \frac{\partial}{\partial y} f(x)(x - z) + Q(e,\eta) \qquad (11)$$

where $Q(e,\eta) = (Q_1(e,\eta), Q_2(e,\eta), Q_3(e,\eta))^T \in R^3$ satisfies

$$Q_i(e,\eta) = \frac{1}{2} e^T G_i(\eta) e \in R, \ i = 1,2,3,$$

$$G_i(\eta) = \begin{pmatrix} \frac{\partial^2}{\partial^2 y_1} & \frac{\partial^2}{\partial y_1 \partial y_2} & \frac{\partial^2}{\partial y_1 \partial y_3} \\ \frac{\partial^2}{\partial y_2 \partial y_1} & \frac{\partial^2}{\partial^2 y_2} & \frac{\partial^2}{\partial y_2 \partial y_3} \\ \frac{\partial^2}{\partial y_3 \partial y_1} & \frac{\partial^2}{\partial y_3 \partial y_2} & \frac{\partial^2}{\partial^2 y_3} \end{pmatrix} f_i(\eta), \ i = 1,2,3,$$

in which $\eta \in R^3$ and

$$\eta = (\eta_1(t), \eta_2(t), \eta_3(t))^T = x + diag\{\theta_1, \theta_2, \theta_3\}(z - x)$$

with $0 < \theta_i < 1 \ (i = 1,2,3)$.

In particular, for the function $f(y)$ given in (1), one has

$$G_1(\eta) = (0)_{3\times 3}, \quad G_2(\eta) = \begin{pmatrix} 0 & 0 & -1 \\ 0 & 0 & 0 \\ -1 & 0 & 0 \end{pmatrix}, \quad G_3(\eta) = \begin{pmatrix} 0 & 1 & 0 \\ 1 & 0 & 0 \\ 0 & 0 & 0 \end{pmatrix}.$$

Thus,

$$Q = (Q_1, Q_2, Q_3)^T = (0, -e_1 e_3, e_1 e_2)^T.$$

By (11), the error system (7) can be reformulated as

$$\dot{e} = \left( A - K + \frac{\partial}{\partial y} f(x) \right) e - Q(e,\eta),$$

for which

$$\| Q \| = (Q_1^2 + Q_2^2 + Q_3^2)^{1/2} \le | e_1 | \| e \|$$

and $\lim_{\|e\|\to 0}\sup \frac{\|Q\|}{\|e\|}=0$.

Hence, it follows from the well-known linearization method for the stability of non-autonomous systems (see Theorem 4.2 of [15]) that if the following linearized system

$$\dot{e}=\left(A-K+\frac{\partial}{\partial y}f(x)\right)e \tag{12}$$

is uniformly asymptotically stable at $e=0$, then the original error system (7) is also uniformly asymptotically stable at $e=0$.

To analyze the stability of system (12), use the quadratic Lyapunov function

$$V(e)=e^{T}Pe,\ 0<P=P^{T}\in R^{3\times 3}.$$

The derivative of $V(e)$ with respect to time $t$ along the trajectory of system (12) equals

$$\dot{V}(e)=\dot{e}^{T}Pe+e^{T}P\dot{e}=e^{T}\left[\left(A-K+\frac{\partial}{\partial y}f(x)\right)^{T}P+P\left(A-K+\frac{\partial}{\partial y}f(x)\right)\right]e.$$

Thus, $\dot{V}(e)<0$ under the condition that the matrix $Y$ defined by (10) is negative definite. This means that the linear time-varying system (12) is uniformly asymptotically stable at the origin (see Theorem 4.1 of [15]), and the proof is completed. ■

Now, consider a special linear controller characterized by the coupling matrix

$$K=\begin{pmatrix} k_{11} & k_{12} & 0 \\ k_{21} & k_{22} & 0 \\ 0 & 0 & k_{33} \end{pmatrix}. \tag{13}$$

With the bounds $\underline{\rho}_i$ and $\overline{\rho}_i$ of the trajectory of master system, let

$$\rho_2=\max\{|\underline{\rho}_2|,|\bar{\rho}_2|\},\ \rho_3=\max\{|\underline{\rho}_3|,|\bar{\rho}_3|\}, \tag{14}$$

$$c_1(p,k_{12},k_{21})=\frac{\rho_3^2-\underline{\rho}_3^2+(\underline{\rho}_3+k_{21}+pk_{12}-a_{21}-pa_{12})^2}{4p}, \tag{15}$$

$$c_2(p,k_{12},k_{21})=\frac{\rho_3^2-\bar{\rho}_3^2+(\bar{\rho}_3+k_{21}+pk_{12}-a_{21}-pa_{12})^2}{4p}, \tag{16}$$

$$c(p,k_{12},k_{21})=\begin{cases}c_1(p,k_{12},k_{21}), & k_{21}+pk_{12}-a_{21}-pa_{12}\le 0,\\ c_2(p,k_{12},k_{21}), & k_{21}+pk_{12}-a_{21}-pa_{12}>0,\end{cases} \tag{17}$$

where the constant $p>0$.

The following theorem gives an algebraic synchronization criterion for the coupling matrix (13).

**Theorem 2.** The master-slave synchronization scheme (6) achieves uniform chaos synchronization if there exists a constant $p>0$ and a coupling matrix $K$ defined by (13) such that

$$k_{11}-a_{11}>0,\ k_{22}-a_{22}>0,\ k_{33}-a_{33}>0, \tag{18}$$

$$(k_{33}-a_{33})\Delta-\frac{1}{4p}(k_{22}-a_{22})\rho_2^2>0, \tag{19}$$

where

$$\Delta=(k_{11}-a_{11})(k_{22}-a_{22})-c(p,k_{12},k_{21}). \tag{20}$$

*Proof.* Take a symmetric positive definite matrix $P=diag\{p_1,p_2,p_2\}$ with $p_i>0\ (i=1,2)$. For the coupling matrix (13), the matrix $Y$ defined by (10) can be represented as

$$Y=\begin{pmatrix}2p_1(a_{11}-k_{11}) & p_1(a_{12}-k_{12})+p_2(a_{21}-k_{21}-x_3) & p_2x_2\\ p_1(a_{12}-k_{12})+p_2(a_{21}-k_{21}-x_3) & 2p_2(a_{22}-k_{22}) & 0\\ p_2x_2 & 0 & 2p_2(a_{33}-k_{33})\end{pmatrix}.$$

Since the matrix $Y$ is symmetric, $Y$ is negative definite if and only if

$$k_{11}-a_{11}>0,\ k_{22}-a_{22}>0,\ k_{33}-a_{33}>0,$$

$$4p_1p_2(k_{11}-a_{11})(k_{22}-a_{22})-[p_1(a_{12}-k_{12})+p_2(a_{21}-k_{21}-x_3)]^2>0,$$

$$8p_1p_2^2(k_{11}-a_{11})(k_{22}-a_{22})(k_{33}-a_{33})-2p_2^2x_2^2(k_{22}-a_{22})$$

$$-2p_2(k_{33}-a_{33})[p_1(a_{12}-k_{12})+p_2(a_{21}-k_{21}-x_3)]^2>0.$$

Let $p=p_1/p_2>0$. Then, the above inequalities are equivalent to

$$k_{11}-a_{11}>0,\ k_{22}-a_{22}>0,\ k_{33}-a_{33}>0, \tag{21}$$

$$(k_{11}-a_{11})(k_{22}-a_{22})-\frac{(x_3+k_{21}+pk_{12}-a_{21}-pa_{12})^2}{4p}>0, \tag{22}$$

$$(k_{33}-a_{33})\left[(k_{11}-a_{11})(k_{22}-a_{22})-\frac{(x_3+k_{21}+pk_{12}-a_{21}-pa_{12})^2}{4p}\right]-\frac{(k_{22}-a_{22})}{4p}x_2^2>0. \tag{23}$$

If $k_{12},k_{21}$ and $p>0$ are selected such that

$$k_{21}+pk_{12}-a_{21}-pa_{12}\le 0,$$

then, for any $t\ge 0$,

$$(x_3+k_{21}+pk_{12}-a_{21}-pa_{12})^2\le \rho_3^2-\underline{\rho}_3^2+(\underline{\rho}_3+k_{21}+pk_{12}-a_{21}-pa_{12})^2.$$

Thus, the inequalities (22) and (23) hold if (18) and (19) are satisfied for $c(p,k_{12},k_{21})=c_1(p,k_{12},k_{21})$.

Similarly, if $k_{12},k_{21}$ and $p$ are selected such that

$$k_{21}+pk_{12}-a_{21}-pa_{12}>0,$$

then, for any $t\ge 0$,

$$(x_3+k_{21}+pk_{12}-a_{21}-pa_{12})^2\le \rho_3^2-\bar{\rho}_3^2+(\bar{\rho}_3+k_{21}+pk_{12}-a_{21}-pa_{12})^2.$$

Thus, inequalities (22) and (23) hold if conditions (18) and (19) are satisfied for $c(p,k_{12},k_{21})=c_2(p,k_{12},k_{21})$.

The proof is completed. ■

From Theorem 2, one can easily obtain a sufficient synchronization criterion for the coupling matrix $K = diag\{k_{11}, k_{22}, k_{33}\}$. Let

$$c_1(p) = c_1(p,0,0) = \frac{\rho_3^2 - \underline{\rho}_3^2 + (\underline{\rho}_3 - a_{21} - pa_{12})^2}{4p}, \tag{24}$$

$$c_2(p) = c_2(p,0,0) = \frac{\rho_3^2 - \bar{\rho}_3^2 + (\bar{\rho}_3 - a_{21} - pa_{12})^2}{4p}, \tag{25}$$

$$c(p) = \begin{cases} c_1(p), & if \ a_{21} - pa_{12} \geq 0, \\ c_2(p), & if \ a_{21} - pa_{12} < 0, \end{cases} \tag{26}$$

with $p > 0$.

**Theorem 3.** The master-slave synchronization scheme (6) achieves uniform chaos synchronization if there exist a constant $p > 0$ and a coupling matrix $K = diag\{k_{11}, k_{22}, k_{33}\}$ such that

$$k_{11} > a_{11}, \tag{27}$$

$$k_{22} > a_{22} + \frac{c(p)}{k_{11} - a_{11}}, \tag{28}$$

$$k_{33} > a_{33} + \frac{\rho_2^2 (k_{22} - a_{22})}{4\bar{\Delta} p}, \tag{29}$$

where

$$\bar{\Delta} = (k_{11} - a_{11})(k_{22} - a_{22}) - c(p). \tag{30}$$

*Proof.* Inequalities (27)-(29) can be immediately derived from (18)-(20) according to the $c(p)$ defined by (24)-(26). ■

***Remark 1.*** The synchronization criteria obtained above are only sufficient but not necessary conditions. Theoretically and practically, it is always expected that the synchronization criterion can provide a wide range of the coupling coefficients, as close as possible to the threshold determined by the necessary synchronization conditions (if exist). Such a criterion is referred to as being *sharp*. Thus, a measure for the quality of a sufficient synchronization criterion is the range of the coupling coefficients produced by the criterion.

Return to criterion (27)-(29), which can be characterized as $k_{ii} > k_{ii}^*(p),\ i = 1,2,3$, in which $k_{ii}^*(p)$ are called *the analytical critical values*. To improve the quality of the criterion, one should select a suitable $p > 0$ such that the critical values are as small as possible.

It follows from (27) and (28) that $\overline{\Delta} > 0$. Thus, minimizing $c(p)$ for $p \in (0,\infty)$ is a reasonable choice for the optimization of the above criterion.

**Lemma 1.** $c(p)$ is a positive continuous function for $p \in (0,\infty)$ and $a_{12} \neq 0$.

*Proof.* From definitions (24)-(26), it is easy to see that $c(p)$ is positive for $p > 0$. Now, assume that $a_{12} > 0$. Then, for $a_{21} \geq 0$, one has $a_{21} + pa_{12} > 0$ for $p \in (0,\infty)$, and

$$c(p) = c_1(p), \quad p \in (0,\infty),$$

which is continuous.

If $a_{12} > 0$ and $a_{21} < 0$, then let $p_0 \in (0,\infty)$ such that

$$a_{21} + p_0 a_{12} = 0. \tag{31}$$

One has

$$\lim_{p \to p_0^+} c(p) = \lim_{p \to p_0} c_1(p) = \frac{1}{4p_0}\rho_3^2,$$

$$\lim_{p \to p_0^-} c(p) = \lim_{p \to p_0} c_2(p) = \frac{1}{4p_0}\rho_3^2.$$

Thus, $c(p)$ is also continuous in this case.

The same method can be applied to prove the result for the case of $a_{12} < 0$. ■

**Lemma 2.** Assume that the (master) system has $a_{12} > 0$. Then, the minimum $c_m$ of $c(p)$ for $p \in (0,\infty)$ equals

$$c_m = \begin{cases} c_1(p_1^*), & \text{if } a_{21} \geq 0, \text{ or } a_{21} < 0 \text{ and } \rho_3^2 - 2a_{21}\underline{\rho}_3 \geq 0, \\ c_2(p_2^*), & \text{if } a_{21} < 0 \text{ and } \rho_3^2 - 2a_{21}\bar{\rho}_3 < 0, \\ c_1(p_0), & \text{else}, \end{cases} \tag{32}$$

where

$$p_1^* = \frac{1}{a_{12}}\sqrt{\rho_3^2 - \underline{\rho}_3^2 + (\underline{\rho}_3 - 2a_{21})^2} > 0, \tag{33}$$

$$p_2^* = \frac{1}{a_{12}}\sqrt{\rho_3^2 - \bar{\rho}_3^2 + (\bar{\rho}_3 - 2a_{21})^2} > 0, \tag{34}$$

and $p_0$ is defined by (31).

*Proof.* For the $c_1(p)$ defined by (24), one has

$$c_1^{'}(p) = -\frac{a_{12}^2}{4p^2}(p^2 - p_1^{*2}),$$

$$c_1^{''}(p) = -\frac{a_{12}^2}{2p^3}(p^2 - p_1^{*2}) + \frac{a_{12}^2}{2p}.$$

Thus, $p_1^*$ is a unique minimal point of $c_1(p)$ for $p \in (0,\infty)$. This means that $c_1(p)$ is decreasing for $p \in (0, p_1^*)$ and increasing for $p \in (p_1^*,\infty)$.

Similarly, one can verify that $p_2^*$ is a unique minimal point of $c_2(p)$ for $p \in (0,\infty)$.

For $a_{12} > 0$, it is clear that

$$p_2^* \begin{cases} > p_1^*, & \text{if } a_{21} < 0, \\ < p_1^*, & \text{if } a_{21} > 0, \\ = p_1^*, & \text{if } a_{21} = 0. \end{cases}$$

Now, analyze the minimal $c_m$ of $c(p)$ for $p \in (0,\infty)$.

If $a_{21} \geq 0$, then $a_{21} + pa_{12} > 0$ for any $p > 0$. Thus, $c(p) = c_1(p)$ for $p \in (0,\infty)$ and $c_m = c_1(p_1^*)$.

If $a_{21} < 0$, then $c(p) = c_2(p)$ for $p \in (0, p_0]$ and $c(p) = c_1(p)$ for $p \in (p_0, \infty)$. Hence, $p_2^* > p_1^* \geq p_0 > 0$ provided that $\rho_3^2 - 2a_{21}\underline{\rho}_3 \geq 0$. This means that for $p \in (0, p_0]$, one has $c(p) = c_2(p)$ which is decreasing; for $p \in (p_0, p_1^*]$, one has $c(p) = c_1(p)$ which is decreasing; and for $p \in (p_1^*, \infty)$, one has $c(p) = c_1(p)$ which is increasing. Thus, $p_1^*$ is a unique minimal point of $c(p)$ for $p \in (0,\infty)$, and $c_m = c_1(p_1^*)$ in this case.

If $a_{21} < 0$ and $\rho_3^2 - 2a_{21}\bar{\rho}_3 \leq 0$, then $p_0 \geq p_2^* > p_1^* > 0$. In this case, for $p \in (0, p_2^*)$, one has $c(p) = c_2(p)$ which is decreasing; for $p \in (p_2^*, p_0]$, one has $c(p) = c_2(p)$ which is increasing; and for $p \in (p_0, \infty)$, one has $c(p) = c_1(p)$ which is increasing. Thus, $p_2^*$ is a unique minimal point of $c(p)$ for $p \in (0,\infty)$, and $c_m = c_2(p_2^*)$.

For the case where $a_{21} < 0$ and $\rho_3^2 - 2a_{21}\bar{\rho}_3 > 0$ and $\rho_3^2 - 2a_{21}\underline{\rho}_3 < 0$, one has $p_2^* > p_0 > p_1^* > 0$. Thus, for $p \in (0, p_0]$, one has $c(p) = c_2(p)$ which is decreasing; and for $p \in (p_0, \infty)$, one has $c(p) = c_1(p)$ which is increasing. Hence, $p_0$ is a unique minimal point of $c(p)$ for $p \in (0,\infty)$ and $c_m = c_1(p_0)$.

The proof is thus completed. ■

Form (24) and (25), it follows that

$$c_1(p_1^*) = \frac{a_{12}}{2}(a_{12}p_1^* + a_{21} - \underline{\rho}_3), \tag{35}$$

$$c_2(p_2^*) = \frac{a_{12}}{2}(a_{12}p_2^* + a_{21} - \bar{\rho}_3), \tag{36}$$

$$c_1(p_0) = -\frac{a_{12}}{4a_{21}}\rho_3^2 \quad with\ a_{21} < 0. \tag{37}$$

According to Lemma 2, one can obtain an optimized synchronization criterion as follows.

**Theorem 4.** The master-slave synchronization scheme (6) with $a_{12} > 0$ achieves uniform chaos synchronization if there exists a coupling matrix $K = diag\{k_{11}, k_{22}, k_{33}\}$ such that

$$k_{11} > a_{11}, \tag{38}$$

$$k_{22} > a_{22} + \frac{c_m}{k_{11} - a_{11}}, \tag{39}$$

$$k_{33} > a_{33} + \frac{\rho_2^2(k_{22} - a_{22})}{4\Delta_m p^*}, \tag{40}$$

where

$$\Delta_m = (k_{11} - a_{11})(k_{22} - a_{22}) - c_m \tag{41}$$

and $c_m$ is defined by (32), with

$$p^* = \begin{cases} p_1^*, & if\ a_{21} > 0,\ or\ a_{21} < 0\ and\ \rho_3^2 - 2a_{21}\underline{\rho}_3 \geq 0, \\ p_2^*, & if\ a_{21} < 0\ \ and\ \rho_3^2 - 2a_{21}\bar{\rho}_3 \leq 0, \\ p_0 & otherwise \end{cases} \tag{42}$$

***Remark 2***. The concerned chaotic systems that satisfy $a_{12} > 0$ conclude the classical Lorenz system, Chen system, Lü system, and the unified chaotic system.

Hence, Theorem 4 covers a broad variety of special cases of the generalized Lorenz systems.

Theorem 4 suggests a simple procedure for designing the coupling coefficients $k_{ii}$ $(i = 1,2,3)$ to achieve chaos synchronization, as follows:

(i) choose $k_{11}$ such that $k_{11} - a_{11} > 0$;

(ii) choose $k_{22}$ such that

$$k_{22} > a_{22} + \frac{c_m}{k_{11} - a_{11}};$$

(iii) choose $k_{33}$ such that

$$k_{33} > a_{33} + \frac{\rho_2^2 (k_{22} - a_{22})}{4\Delta_m p^*},$$

where $\Delta_m$ and $p^*$ are defined by (41) and (42), respectively.

The following results present some chaos synchronization criteria for single variable-coupled configurations, which correspond to the simplest possible linear synchronization controllers.

**Theorem 5.** The master-slave synchronization scheme (6) with $a_{12} > 0$ and $a_{21} > 0$ and $a_{ii} < 0$ $(i = 2,3)$ achieves uniform chaos synchronization if there exist a constant $p > 0$ and a coupling matrix $K = diag\{k_{11}, 0, 0\}$ such that

$$k_{11} > a_{11} + \frac{1}{|a_{22}|} c_1(p) + \frac{1}{4p|a_{33}|} \rho_2^2, \tag{43}$$

where $c_1(p)$ is defined by (24).

*Proof.* For the considered scheme, one has

$$a_{21} + pa_{12} > 0, \quad p \in (0, \infty).$$

Thus, $c(p) \equiv c_1(p)$.

The synchronization conditions (27)-(29) can be satisfied provided that

$$k_{11} - a_{11} > 0, \text{ and}$$

$$k_{11} > a_{11} + \frac{1}{|a_{22}|} c_1(p) + \frac{1}{4p|a_{33}|} \rho_2^2.$$

Clearly, condition (43) guarantees the above inequalities to hold. ■

Let

$$G(p) = \frac{1}{|a_{22}|} c_1(p) + \frac{1}{4p|a_{33}|} \rho_2^2. \tag{44}$$

Finding a $p > 0$ such that $G(p)$ achieves its minimum will improve the sharpness of the synchronization criterion (43). This expectation is realized by the following lemma and theorem.

**Lemma 3.** Assume $a_{12} > 0$. Then, $G(p)$ achieves the following minimum $G_m$ for $p \in (0, \infty)$:

$$G_m = G(p_3^*) = \frac{a_{12}}{2|a_{22}|} (a_{12} p_3^* + a_{21} - \underline{\rho}_3) \tag{45}$$

at the unique minimal point

$$p_3^* = \frac{1}{a_{12}} \sqrt{\rho_3^2 - \underline{\rho}_3^2 + (\underline{\rho}_3 - a_{21})^2 + \frac{|a_{22}|}{|a_{33}|} \rho_2^2} > 0. \tag{46}$$

*Proof.* It is easy to obtain the minimal point $p_3^*$ and the minimum value $G_m$ of $G(p)$ by calculating the first-order and second-order derivatives of $G(p)$ for $p$. ■

**Theorem 6.** The master-slave synchronization scheme (6) with $a_{12} > 0$ and $a_{21} > 0$ and $a_{ii} < 0$ $(i = 2,3)$ achieves uniform chaos synchronization if there exists a coupling matrix $K = diag\{k_{11}, 0, 0\}$ such that

$$k_{11} > a_{11} + G_m, \tag{47}$$

where $G_m$ is determined by (45) and (46).

***Remark 3.*** Theorem 5 and Theorem 6 are applicable to a class of generalized Lorenz systems containing the classical Lorenz system and the unified chaotic system with $0 \le \alpha \le 1/29$.

**Theorem 7.** The master-slave synchronization scheme (6) with $a_{ii} < 0$ $(i = 1, 3)$ achieves uniform chaos synchronization if there exist a constant $p > 0$ and a coupling matrix $K = diag\{0, k_{22}, 0\}$ such that

$$4pa_{11}a_{33} - \rho_2^2 > 0, \tag{48}$$

$$k_{22} > a_{22} + \frac{4p\,|a_{33}|}{4pa_{11}a_{33} - \rho_2^2} c(p), \tag{49}$$

where $c(p)$ is defined by (24)-(26).

*Proof.* According to (48), one has

$$\frac{4p\,|a_{33}|}{4pa_{11}a_{33} - \rho_2^2} > \frac{1}{|a_{11}|} > 0.$$

Thus, the synchronization conditions (27)-(29) hold for $k_{11} = k_{33} = 0$, if $a_{11} < 0$ and $a_{33} < 0$, and inequality (49) is satisfied. ■

Now, consider the function

$$L(p) = \frac{4p\,|a_{33}|}{4pa_{11}a_{33} - \rho_2^2} c(p) \tag{50}$$

with $a_{11} < 0$ and $a_{33} < 0$. Let $p_k \in R^+$ and satisfy

$$4p_k a_{11}a_{33} - \rho_2^2 = 0. \tag{51}$$

In order to optimize the synchronization criterion (49), one has to minimize the function $L(p)$ for $p \in (0, \infty)$ under condition (48), or equivalently, to minimize $L(p)$ for $p \in (p_k, \infty)$.

Define

$$L_1(p) = \frac{4p|a_{33}|}{4pa_{11}a_{33} - \rho_2^2} c_1(p), \quad p \in (0, \infty), \tag{52}$$

$$L_2(p) = \frac{4p|a_{33}|}{4pa_{11}a_{33} - \rho_2^2} c_2(p), \quad p \in (0, \infty), \tag{53}$$

$$p_4^* = \frac{\rho_2^2}{4a_{11}a_{33}} + \sqrt{\left(\frac{\rho_2^2}{4a_{11}a_{33}} - \frac{\underline{\rho}_3 - a_{21}}{a_{12}}\right)^2 + \frac{\rho_3^2 - \underline{\rho}_3^2}{a_{12}^2}}, \tag{54}$$

$$p_5^* = \frac{\rho_2^2}{4a_{11}a_{33}} + \sqrt{\left(\frac{\rho_2^2}{4a_{11}a_{33}} - \frac{\bar{\rho}_3 - a_{21}}{a_{12}}\right)^2 + \frac{\rho_3^2 - \bar{\rho}_3^2}{a_{12}^2}}. \tag{55}$$

**Lemma 4.** Assume that the chaotic (master) system has $a_{12} > 0$ and $a_{ii} < 0$ $(i = 1, 3)$. Let

$$\Omega = \frac{a_{21}}{a_{12}} + \frac{\rho_2^2}{4a_{11}a_{33}}. \tag{56}$$

Then, the minimum of $L(p)$ for $p \in (p_k, \infty)$ equals

$$L_m = \begin{cases} L_1(p_4^*), \; \textit{if } \Omega \geq 0, \; \textit{or } \Omega < 0 \; \textit{and } a_{21} + p_4^* a_{12} \geq 0, \\ L_2(p_5^*), \; \textit{if } \Omega < 0 \; \textit{and } a_{21} + p_5^* a_{12} \leq 0, \\ L_1(p_0), \; \textit{otherwise}, \end{cases} \tag{57}$$

where $p_4^*$, $p_5^*$ and $p_0$ are defined by (54), (55) and (31), respectively.

*Proof.* By some complex algebraic operations, one can obtain the derivatives of $L_1(p)$ for $p$, as

$$L_1'(p) = \sigma_1(p)\left[\left(p - \frac{\rho_2^2}{4a_{11}a_{33}}\right)^2 - \left(\frac{\rho_2^2}{4a_{11}a_{33}} - \frac{\underline{\rho}_3 - a_{21}}{a_{12}}\right)^2 - \frac{\rho_3^2 - \underline{\rho}_3^2}{a_{12}^2}\right],$$

$$L_1''(p) = -\sigma_2(p)\frac{L_1'(p)}{\sigma_1(p)} + 2\sigma_1(p)\left(p - \frac{\rho_2^2}{4a_{11}a_{33}}\right),$$

where

$$\sigma_1(p) = \frac{4a_{12}^2 a_{33}^2 \mid a_{11} \mid}{(4pa_{11}a_{33} - \rho_2^2)^2},$$

$$\sigma_2(p) = \frac{2\sigma_1(p)}{4pa_{11}a_{33} - \rho_2^2}.$$

Thus, $p_4^*$ defined by (54) is a unique minimal point of $L_1(p)$.

Again, $4p_4^* a_{11}a_{33} - \rho_2^2 > 0$. This means that $p_4^* > p_k > 0$. Hence, $L_1(p)$ is decreasing for $p \in (p_k, p_4^*)$ and increasing for $p \in (p_4^*, \infty)$.

Similarly, it can be verified that $p_5^*$, defined by (55), is a unique minimal point of $L_2(p)$ and $p_5^* > p_k > 0$. Thus, $L_2(p)$ is decreasing for $p \in (p_k, p_5^*)$ and increasing for $p \in (p_5^*, \infty)$.

Apparently,

$$p_4^* \begin{cases} > p_5^*, & if\, \Omega > 0, \\ = p_5^*, & if\, \Omega = 0, \\ < p_5^*, & if\, \Omega < 0, \end{cases}$$

where $\Omega$ is defined by (56).

Next, analyze the minimum $L_m$ of $L(p)$ for $p \in (p_k, \infty)$.

It is already known that $c(p)$ is a continuous positive function for $p \in (p_k, \infty)$ and so is $L(p)$.

If $a_{21} \geq 0$, then $\Omega > 0$ and $a_{21} + pa_{12} > 0$ for $p \in (p_k, \infty)$. Thus, $L(p) = L_1(p)$ for $p \in (p_k, \infty)$, and the minimum is $L_m = L_1(p_4^*)$.

If $a_{21} < 0$ and $\Omega \geq 0$, then $p_4^* \geq p_5^* > p_k \geq p_0$. This shows that $L(p) = L_1(p)$ for $p \in (p_k, \infty)$. Thus, we have $L_m = L_1(p_4^*)$.

If $a_{21} < 0$ and $\Omega < 0$, then $p_0 > p_k$ and $p_5^* > p_4^*$. For this case, one must classify the parameters $(a_{21}, a_{12})$ in order to solve for the minimum $L_m$.

The first case is $a_{21} + p_4^* a_{12} \geq 0$. In this case, $p_4^* > p_0$, so one has $p_5^* > p_4^* > p_0 > p_k$. Then, for $p \in (p_k, p_0]$, one has $L(p) = L_2(p)$ which is decreasing; for $p \in (p_0, p_4^*)$, one has $L(p) = L_1(p)$ which is decreasing; for $p \in (p_4^*, \infty)$, one has $L(p) = L_1(p)$ which is increasing. Hence, $p_4^*$ is a unique minimal point of $L(p)$ for $p \in (p_k, \infty)$, and $L_m = L_1(p_4^*)$.

The next case is $a_{21} + p_5^* a_{12} \leq 0$, which leads to $p_5^* \leq p_0$. So, one has $p_0 \geq p_5^* > p_k$. Then, for $p \in (p_k, p_5^*]$, one has $L(p) = L_2(p)$ which is decreasing; for $p \in (p_5^*, p_0)$, one has $L(p) = L_2(p)$ which is increasing; for $p \in [p_0, \infty)$, one has $L(p) = L_1(p)$ which is increasing because of $p_4^* < p_5^* \leq p_0$ in this case. Hence, $p_5^*$ is a unique minimal point of $L(p)$ for $p \in (p_k, \infty)$, and $L_m = L_2(p_5^*)$.

Finally, consider the case of $a_{21} + p_4^* a_{12} < 0$ and $a_{21} + p_5^* a_{12} > 0$. In this case, $p_4^* < p_0 < p_5^*$. Thus, for $p \in (p_k, p_0)$, one has $L(p) = L_2(p)$ which is decreasing; for $p \in [p_0, \infty)$, one has $L(p) = L_1(p)$ which is increasing. Hence, $p_0$ is unique minimal point of $L(p)$ for $p \in (p_k, \infty)$, and $L_m = L_1(p_0)$.

The proof is completed. ■

**Theorem 8.** The master-slave synchronization scheme (6) with $a_{12} > 0$ and $a_{ii} < 0$ $(i = 1,3)$ achieves uniform chaos synchronization if there exists a coupling matrix $K = diag\{0, k_{22}, 0\}$ such that

$$k_{22} > a_{22} + L_m, \tag{58}$$

where $L_m$ is defined by (57).

***Remark 4.*** There exist some special types of generalized Lorenz systems satisfying the conditions $a_{12}>0$ and $a_{ii}<0$ $(i=1,3)$, including the classical Lorenz system, Chen system, Lü system and the unified chaotic system.

For convenience in calculating $L_m$, some relative formulas are listed below:

$$L_1(p_4^*)=\frac{a_{12}^2}{2|a_{11}|}\left(p_4^*-\frac{\underline{\rho}_3-a_{21}}{a_{12}}\right), \tag{59}$$

$$L_2(p_5^*)=\frac{a_{12}^2}{2|a_{11}|}\left(p_5^*-\frac{\rho_3^2-\overline{\rho}_3^2}{a_{12}^2}\right), \tag{60}$$

$$L_1(p_0)=\frac{|a_{33}|a_{12}\rho_3^2}{4|a_{21}|a_{11}a_{33}-\rho_2^2}, \tag{61}$$

with $a_{12}>0$, $a_{11}<0$ and $a_{33}<0$.

From Theorem 3, one can easily prove the following synchronization criterion for the coupling matrix $K=diag\{0,0,k_{33}\}$.

**Theorem 9.** The master-slave synchronization scheme (6) with $a_{12}>0$, $a_{21}>0$ and $a_{ii}<0$ $(i=1,2)$ achieves uniform chaos synchronization if there exists a constant $p>0$ and a coupling matrix $K=diag\{0,0,k_{33}\}$ such that

$$c_1(p)-a_{11}a_{22}<0, \tag{62}$$

$$k_{33}>a_{33}+\frac{|a_{22}|}{4p(a_{11}a_{21}-c_1(p))}\rho_2^2, \tag{63}$$

where $c_1(p)$ is defined by (24).

Let

$$Q(p)=4p(a_{11}a_{22}-c_1(p)) \tag{64}$$

with $a_{11}<0$ and $a_{22}<0$.

It is intended to maximize $Q(p)$ for $p \in (0,\infty)$ under condition (62), so as to improve the synchronization condition (63).

**Lemma 5.** Inequality (62) has the solution of $p > 0$ if and only if

$$\frac{2a_{11}a_{22}}{a_{12}} + \underline{\rho}_3 - a_{21} > \sqrt{(\underline{\rho}_3 - a_{21})^2 + \rho_3^2 - \underline{\rho}_3^2}, \tag{65}$$

and its solution satisfies

$$\bar{p} > p > \underline{p} > 0, \tag{66}$$

where

$$\bar{p} = \frac{1}{a_{12}}\left[\left(\frac{2a_{11}a_{22}}{a_{12}} + \underline{\rho}_3 - a_{21}\right) + \sqrt{\left(\frac{2a_{11}a_{22}}{a_{12}} + \underline{\rho}_3 - a_{21}\right)^2 - (\underline{\rho}_3 - a_{21})^2 - \rho_3^2 + \underline{\rho}_3^2}\right], \tag{67}$$

$$\underline{p} = \frac{1}{a_{12}}\left[\left(\frac{2a_{11}a_{22}}{a_{12}} + \underline{\rho}_3 - a_{21}\right) - \sqrt{\left(\frac{2a_{11}a_{22}}{a_{12}} + \underline{\rho}_3 - a_{21}\right)^2 - (\underline{\rho}_3 - a_{21})^2 - \rho_3^2 + \underline{\rho}_3^2}\right]. \tag{68}$$

According to this lemma, the optimization issue can be described as maximizing $Q(p)$ for $p \in (\underline{p}, \bar{p})$. The optimized result is presented in the following lemma.

**Lemma 6.** Assume that the master chaotic system has $a_{12} > 0$, $a_{21} > 0$ and $a_{ii} < 0$ $(i = 1, 2)$, and satisfies inequality (65). Then, for $p \in (\underline{p}, \bar{p})$, $Q(p)$ achieves the maximum

$$Q_m = Q(p_6^*) = 4a_{11}a_{22}\left(\frac{a_{11}a_{22}}{a_{12}^2} + p_6^*\right) - \rho_3^2 + \underline{\rho}_3^2, \tag{69}$$

at the unique maximal point

$$p_6^* = \frac{1}{a_{12}}\left(\frac{2a_{11}a_{22}}{a_{12}} + \underline{\rho}_3 - a_{21}\right) \in (\underline{p}, \bar{p}). \tag{70}$$

**Theorem 10.** The master-slave synchronization scheme (6) with $a_{12} > 0$, $a_{21} > 0$ and $a_{ii} < 0$ $(i = 1,2)$ achieves uniform chaos synchronization if the master system satisfies inequality (65) and there exists a coupling matrix $K = diag\{0,0,k_{33}\}$ such that

$$k_{33} > a_{33} + \frac{\rho_2^2 \,|\, a_{22} \,|}{Q_m}, \tag{71}$$

where $Q_m$ is determined by (69) and (70).

***Remark 5.*** The specific generalized Lorenz systems that satisfy $a_{12} > 0$, $a_{21} > 0$ and $a_{ii} < 0$ $(i = 1, 2)$ are the classical Lorenz system and the unified chaotic system with $0 \le \alpha \le 1/29$. Theorem 9 and Theorem 10 are applicable to these types of chaotic systems.

Now, consider the coupling matrix

$$K = \begin{pmatrix} 0 & k_{12} & 0 \\ 0 & 0 & 0 \\ 0 & 0 & 0 \end{pmatrix}, \tag{72}$$

which results in a so-called *dislocated feedback control* applicable to the synchronization scheme studied in [7].

Let

$$c_1(p, k_{12}) = \frac{\rho_3^2 - \underline{\rho}_3^2 + (pk_{12} + \underline{\rho}_3^2 - a_{21} - pa_{12})^2}{4p}, \tag{73}$$

$$c_2(p, k_{12}) = \frac{\rho_3^2 - \bar{\rho}_3^2 + (pk_{12} + \bar{\rho}_3 - a_{21} - pa_{12})^2}{4P}, \tag{74}$$

$$c(p, k_{12}) = \begin{cases} c_1(p, k_{12}), \; pk_{12} - a_{21} - pa_{12} \le 0, \\ c_2(p, k_{12}), \; pk_{12} - a_{21} - pa_{12} \ge 0, \end{cases} \tag{75}$$

with $p > 0$.

**Theorem 11.** The master-slave synchronization scheme (6) with $a_{ii} < 0$ $(i = 1, 2, 3)$ achieves uniform chaos synchronization if there exist a constant $p > 0$ and a coupling matrix $K$, defined by (72), such that either of the following synchronization conditions is satisfied:

$$\text{(i)}\quad a_{12} + \frac{1}{p}(a_{21} - \underline{\rho}_3 - \underline{m}(p)) < k_{12} < a_{12} + \frac{1}{p}(a_{21} - \overline{\rho}_3 - \overline{m}(p)), \tag{76}$$

where $p \in R^+$ and

$$p > \frac{1}{4a_{11}a_{22}}\left(\rho_3^2 + \frac{a_{22}}{a_{33}}\rho_2^2\right), \tag{77}$$

$$\overline{m}(p) = \sqrt{4pa_{11}a_{22} - \frac{a_{22}}{a_{33}}\rho_2^2 - \rho_3^2 + \overline{\rho}_3^2}, \tag{78}$$

$$\underline{m}(p) = \sqrt{4pa_{11}a_{22} - \frac{a_{22}}{a_{33}}\rho_2^2 - \rho_3^2 + \underline{\rho}_3^2}\ ; \tag{79}$$

$$\text{(ii)}\quad a_{12} + \frac{1}{p}(a_{21} - \underline{\rho}_3 - \underline{m}(p)) < k_{12} < a_{12} + \frac{1}{p}(a_{21} - \underline{\rho}_3 - \underline{m}(p)) \tag{80}$$

with $\underline{\rho}_3 \geq 0$ and

$$\frac{1}{4a_{11}a_{22}}\left(\frac{a_{22}}{a_{33}}\rho_2^2 + \rho_3^2\right) \geq p \geq \frac{1}{4a_{11}a_{22}}\left(\frac{a_{22}}{a_{33}}\rho_2^2 + \rho_3^2 - \underline{\rho}_3^2\right), \tag{81}$$

with $\overline{m}(p)$ and $\underline{m}(p)$ defined by (78) and (79), respectively.

*Proof.* It is known that $\overline{m}(p)$, $\underline{m}(p) \in R^+$ under conditions (77) and (81).

According to the selection of $K$ and $a_{ii} < 0$ $(i = 1,2,3)$, the synchronization conditions (18) and (19) hold provided that

$$c(p, k_{12}) > a_{11}a_{22} - \frac{a_{22}}{a_{33}}\frac{\rho_2^2}{4p} \tag{82}$$

with $a_{ii} < 0$ $(i = 1,2,3)$.

Inequality (76) has the following equivalent form:

$$-\underline{m}(p) - \underline{\rho}_3 < pk_{12} - a_{21} - pa_{12} < \overline{m}(p) - \overline{\rho}_3 .$$

Again, when inequality (77) is satisfied, one has $\overline{m}(p) > |\overline{\rho}_3| > 0$ and $\underline{m}(p) > |\underline{\rho}_3| > 0$.

Thus, if $-\underline{m}(p) - \underline{\rho}_3 < pk_{12} - a_{21} - pa_{12} \le 0$, then $c(p,k_{12}) = c_1(p,k_{12})$ and the synchronization condition (82) holds; if $0 \le pk_{12} - a_{21} - pa_{12} < \overline{m}(p) - \overline{\rho}_3$, then $c(p,k_{12}) = c_2(p,k_{12})$ and condition (82) also holds. This means that synchronization condition (82) is satisfied if inequalities (76) and (77) hold.

Again, inequality (80) is equivalent to

$$-\underline{m}(p) - \underline{\rho}_3 < pk_{12} - a_{21} - pa_{12} < \underline{m}(p) - \underline{\rho}_3 . \tag{83}$$

Based on $\underline{\rho}_3 \ge 0$ and inequality (81), one has $0 < \underline{m}(p) < \underline{\rho}_3$. Thus, inequality (83) implies that $pk_{12} - a_{21} - pa_{12} < 0$. In this case, $c(p,k_{12}) = c_1(p,k_{12})$ and synchronization condition (82) is satisfied provided that inequalities (80) and (81) hold.

The proof is thus completed. ■

***Remark 6.*** There are several methods for optimizing the synchronization criteria (76)-(77) and (80)-(81), so as to reduce their conservativeness. For example, a method is to design $p$ such that the interval of $k_{12}$ determined by (76) or (80) is maximal under condition (77) or (81). In this way, the objective function for optimization is chosen to be $(\overline{m}(p) + \underline{m}(p) - \overline{\rho}_3 + \underline{\rho}_3)/p$ for criterion (76), or $\underline{m}(p)/p$ for criterion (80). Another method is to select $p$ such that the upper bounds of criterion (76) or (80) are maximal, or their lower bounds are minimal, under the constraint conditions (77) and (81). Then, the objective function must be these bounds. Hence, several optimized synchronization criteria can be produced in light of the selected objective functions.

Consider another "dislocated" coupling matrix

$$K = \begin{pmatrix} 0 & 0 & 0 \\ k_{21} & 0 & 0 \\ 0 & 0 & 0 \end{pmatrix}. \tag{84}$$

Let

$$c_1(p,k_{21}) = \frac{\rho_3^2 - \underline{\rho}_3^2 + (k_{21} + \underline{\rho}_3 - a_{21} - pa_{12})^2}{4p}, \tag{85}$$

$$c_2(p,k_{21}) = \frac{\rho_3^2 - \bar{\rho}_3^2 + (k_{21} + \bar{\rho}_3 - a_{21} - pa_{12})^2}{4p}, \tag{86}$$

$$c\,(p,k_{21}) = \begin{cases} c_1(p,k_{21}), & k_{21} - a_{21} - pa_{12} \le 0, \\ c_2(p,k_{21}), & k_{21} - a_{21} - pa_{12} > 0. \end{cases} \tag{87}$$

**Theorem 12.** The master-slave synchronization scheme (6) with $a_{ii} < 0\,(i = 1,2,3)$ achieves uniform chaos synchronization if there exist a constant $p > 0$ and a coupling matrix $K$, defined by (84), such that either of the following synchronization criteria is satisfied:

(i) $a_{21} + pa_{12} - \underline{\rho}_3 - \underline{m}(p) < k_{21} < a_{21} + pa_{21} - \bar{\rho}_3 + \bar{m}(p)$, (88)

where $p$ satisfies (77), and $\bar{m}(p)$ and $\underline{m}(p)$ are defined by (78) and (79), respectively;

(ii) $a_{21} + pa_{12} - \underline{\rho}_3 - \underline{m}(p) < k_{21} < a_{21} + pa_{12} - \underline{\rho}_3 + \underline{m}(p)$, (89)

where $\underline{\rho}_3 \ge 0$, $p$ satisfies (81), and $\underline{m}(p)$ is defined by (79).

*Proof.* It is similar to the proof of Theorem 11. ■

***Remark** 7.* Similar to the discussion in *Remark 6*, analysis on optimizing criteria (88) and (89) can be given. Besides, since the classical Lorenz system and the unified chaotic system with $\alpha \in [0, 1/29)$ have $a_{ii} < 0\,(i = 1,2,3)$, Theorem 11 and Theorem 12 are applicable to such generalized Lorenz systems.

Owing to the limitation of space, the sufficient synchronization criteria for scheme (6) with the special coupling matrix $K$ defined only by the coupling coefficients $k_{13}, k_{31}, k_{23}$ or $k_{32}$ are not further discussed here.

## 4. Applications and Simulations

In this section, the theoretical results are illustrated with four typical chaotic systems, *i.e.*, the classical Lorenz system, Chen system, Lü system and a unified chaotic system. The newly developed synchronization criteria are verified and then compared with the existing criteria.

### 4.1. The classical Lorenz system

Some sufficient criteria for the synchronization scheme (6) with the Lorenz system were investigated in [6-10].

According to the parameters of the Lorenz system given in (2), one can obtain the new synchronization criteria for the scheme using Theorems 4, 6, 8, 10, 11, and 12, as summarized in the following proposition.

**Proposition 1.** The master-slave synchronization scheme (6) for the classical Lorenz system achieves uniform chaos synchronization if any of the following synchronization criterion is satisfied:

(i) $K = diag\{k_{11}, k_{22}, k_{33}\}$ with

$$\begin{cases} k_{11} + a > 0\,, \\ k_{22} > \dfrac{a}{2(k_{11}+a)}\left(\sqrt{\rho_3^2 + (\underline{\rho}_3 - c)^2 - \underline{\rho}_3^2} + c - \underline{\rho}_3\right) - 1, \\ k_{33} > \dfrac{a\rho_2^2(k_{22}+1)}{4\bar{\Delta}_L\sqrt{\rho_3^2 + (\underline{\rho}_3 - c)^2 - \underline{\rho}_3^2}} - b\,, \end{cases} \tag{90}$$

where

$$\bar{\Delta}_L = (k_{11}+a)(k_{22}+1) - \frac{a}{2}\left(\sqrt{\rho_3^2 + (\underline{\rho}_3 - c)^2 - \underline{\rho}_3^2} + c - \underline{\rho}_3\right); \tag{91}$$

(ii) $K = diag\{k_{11}, 0, 0\}$ with

$$k_{11} > \frac{a}{2}\left(\sqrt{\rho_3^2 + (\underline{\rho}_3 - c)^2 - \underline{\rho}_3^2 + \frac{1}{b}\rho_2^2} + c - \underline{\rho}_3 - 2\right); \tag{92}$$

(iii) $K = diag\{0, k_{22}, 0\}$ with

$$k_{22} > \frac{1}{2}\left(\sqrt{\rho_3^2 + (\frac{1}{4b}\rho_2^2 - \underline{\rho}_3 + c)^2 - \underline{\rho}_3^2} + \frac{1}{4b}\rho_2^2 + c - \underline{\rho}_3 - 2\right); \tag{93}$$

(iv) $K = diag\{0, 0, k_{33}\}$ with

$$2 + \underline{\rho}_3 - c > \sqrt{\rho_3^2 + (\underline{\rho}_3 - c)^2 - \underline{\rho}_3^2} \tag{94}$$

and

$$k_{33} > 4(3 + \underline{\rho}_3 - c) - \rho_3^2 + \underline{\rho}_3^2 - b; \tag{95}$$

(v) $K = \begin{pmatrix} 0 & k_{12} & 0 \\ 0 & 0 & 0 \\ 0 & 0 & 0 \end{pmatrix}$ with

$$a + \frac{1}{p}\left(c - \underline{\rho}_3 - \sqrt{4ap - \frac{1}{b}\rho_2^2 + \underline{\rho}_3^2 - \rho_3^2}\right) < k_{12} < a + \frac{1}{p}\left(c - \bar{\rho}_3 - \sqrt{4ap - \frac{1}{b}\rho_2^2 + \bar{\rho}_3^2 - \rho_3^2}\right), \tag{96}$$

where

$$p > \frac{1}{4a}\left(\frac{1}{b}\rho_2^2 + \rho_3^2\right), \tag{97}$$

or,

$$a+\frac{1}{p}\left(c-\underline{\rho}_3-\sqrt{4ap-\frac{1}{b}\rho_2^2+\underline{\rho}_3^2-\rho_3^2}\right)<k_{12}<a+\frac{1}{p}\left(c-\underline{\rho}_3+\sqrt{4ap-\frac{1}{b}\rho_2^2+\underline{\rho}_3^2-\rho_3^2}\right), \tag{98}$$

where

$$\frac{1}{4a}\left(\frac{1}{b}\rho_2^2+\rho_3^2\right)\geq p\geq\frac{1}{4a}\left(\frac{1}{b}\rho_2^2+\rho_3^2-\underline{\rho}_3^2\right)>0; \tag{99}$$

(vi) $K=\begin{pmatrix} 0 & 0 & 0 \\ k_{21} & 0 & 0 \\ 0 & 0 & 0 \end{pmatrix}$ with

$$c+ap-\underline{\rho}_3-\sqrt{4ap-\frac{1}{b}\rho_2^2+\underline{\rho}_3^2-\rho_3^2}<k_{21}<c+ap-\bar{\rho}_3+\sqrt{4ap-\frac{1}{b}\rho_2^2+\bar{\rho}_3^2-\rho_3^2}, \tag{100}$$

where $p$ satisfies (97), or,

$$c+ap-\underline{\rho}_3-\sqrt{4ap-\frac{1}{b}\rho_2^2+\underline{\rho}_3^2-\rho_3^2}<k_{21}<c+ap-\underline{\rho}_3+\sqrt{4ap-\frac{1}{b}\rho_2^2+\underline{\rho}_3^2-\rho_3^2}, \tag{101}$$

where $p$ satisfies (99).

To the best of our knowledge, the sufficient synchronization criteria for the classical Lorenz systems, linearly coupled by $K=diag\{0,k_{22},0\}$ or $K=diag\{0,0,k_{33}\}$, are not available in the existing literature.

It should be noticed that the bounds of the trajectory of the master chaotic system have to be estimated before any of the above synchronization criteria can be used. The precision of the estimation sensitively influence the sharpness of the synchronization criteria.

For the classical Lorenz system, the trajectory of the chaotic attractor has been proven to satisfy [16]

$$x_1^2+x_2^2+(x_3-a-c)^2\leq R_1^2 \tag{102}$$

where

$$R_1^2 = \begin{cases} \dfrac{(a+c)^2 b^2}{4(b-1)}, & a \ge 1, b \ge 2, \\ \dfrac{(a+c)^2 b^2}{4a(b-a)}, & 0 < a < 1, b \ge 2a, \\ (a+c)^2, & a > \dfrac{b}{2}, 0 < b < 2, \end{cases} \tag{103}$$

or, more precisely [17][18],

$$|x_1| \le R_2, \quad x_2^2 + (x_3 - c)^2 \le R_2^2, \tag{104}$$

where

$$R_2 = \begin{cases} c, & 0 < b \le 2, \\ \dfrac{bc}{2\sqrt{b-1}}, & b > 0. \end{cases} \tag{105}$$

Thus, the following analytical estimations for the bounds of the master Lorenz system can be solved based on (102)-(105) if the initial state of the master chaotic system is selected from inside a neighborhood of its attractor:

$$\text{(i)} \quad \rho_2 = R_1, \underline{\rho}_3 = a + c - R_1, \bar{\rho}_3 = \rho_3 = a + c + R_1, \tag{106}$$

or,

$$\text{(ii)} \quad \rho_2 = R_2, \underline{\rho}_3 = c - R_2, \bar{\rho}_3 = \rho_3 = c + R_2. \tag{107}$$

However, numerical simulations show that the above estimation is very rough. For example, if the typical parameters of the classical Lorenz system, $a = 10, b = 8/3$ and $c = 28$, are chosen, then the lower bound $\underline{\rho}_3$ can be estimated as $-1502.3$ by using formula (106), or $-0.92$ by formula (107). But simulation shows that a more precise estimation is $\underline{\rho}_3 = 6.9$, as seen from Fig.1.

Fig.1

To show the advantage of the new criteria derived above in this paper, a new concept is first introduced.

**Definition 1.** A sufficient synchronization criterion “A” is referred to be sharper than a criterion “B” if the ranges of the coupling coefficients produced by “A” cover the corresponding ranges produced by “B”.

Recall the following form of synchronization criterion for the scheme with the unidirectional coupling $K = diag\{k_{11}, k_{22}, k_{33}\}$ [9] and the bidirectional coupling $K_1 = diag\{d_{11}, d_{12}, d_{13}\}$ and $K_2 = diag\{d_{21}, d_{22}, d_{23}\}$ [8], respectively, obtained by using a Lyapunov function $P = diag\{p_1, p_2, p_2\}$ with $p_i > 0$ $(i = 1,2)$:

$$\begin{cases} r_1 > -a + \dfrac{\varepsilon}{2p_1}, \\ r_2 > \dfrac{(p_1 a + p_2(c - x_3))^2}{4p_1 p_2 (a + r_1) - 2p_2\varepsilon} - 1 + \dfrac{\varepsilon}{2p_2}, \\ r_3 > -b + \dfrac{\varepsilon}{2p_2} + \dfrac{p_2 x_2^2 (-\varepsilon + 2(1 + r_2)p_2)^2}{2\Delta_2}, \end{cases} \tag{108}$$

where $r_i = k_{ii}$ for unidirectional coupling and $r_i = d_{1i} + d_{2i}$ for bidirectional coupling $(i = 1,2,3)$,

$$\Delta_2 = 4p_1 p_2 \left(a + r_1 - \frac{\varepsilon}{2p_1}\right)\left(1 + r_2 - \frac{\varepsilon}{2p_2}\right) - (p_1 a + (c - x_3)p_2)^2 \tag{109}$$

and $\varepsilon \geq 0$ is a small constant.

For the classical Lorenz system, with $a = 16$, $b = 4$ and $c = 45.6$, a synchronization condition is obtained in [9] with

$$K = diag\{45.6, 16, 110\}, \tag{110}$$

by using criterion (108) and letting $p_1 = p_2 = 1$ and $\varepsilon = 0.1$.

From Fig.10 of [9], the master Lorenz system has the following bounds:

$$\rho_2 = 40, \quad \underline{\rho}_3 = 10, \quad \rho_3 = \overline{\rho}_3 = 75. \tag{111}$$

According to the new criterion defined by (90)-(91) and the procedure suggested in Section 3, the following new synchronization conditions can be obtained:

$$\begin{cases} K = diag\{45.6, 16, 12.9\}, \\ K = diag\{45.6, 14.5, 110\}, \\ K = diag\{40.3, 16, 110\}. \end{cases}$$

The corresponding coupling coefficients are smaller than or equal to those coefficients shown in (110). This means that the new criterion defined by (90)-(91) is sharper than the existing criterion (108), at least in this example.

Chaos synchronization performance of the master-slave Lorenz systems coupled by $K = diag\{45.6, 16, 12.9\}$ is illustrated in Fig.2.

Fig.2

In [8], the classical Lorenz system with $a = 10, b = 8/3$ and $c = 28$ is considered, showing that if $p_1 = p_2 = 1$ and $\varepsilon = 0.4$, then the coupling coefficients $d_{11} = d_{21} = 0.5$, $d_{12} = d_{22} = 3$, and $d_{13} = d_{23} = 3$ can satisfy criterion (108) for any trajectory $(x_1, x_2, x_3)$ of the master system.

However, a counterexample can be easily found: with $p_1 = p_2 = 1$, $\varepsilon = 0.4$ and $d_{11} = d_{21} = 0.5$, the synchronization condition derived from (108) is

$$d_{12} + d_{22} > \frac{(38 - x_3)^2}{43.2} - 0.8. \tag{112}$$

Figure 1 shows $6.9 < x_3 < 44.2$. Thus, there must exist a trajectory, $x_3 = 8$, such that condition (112) does not hold for $d_{12} = d_{22} = 3$. Hence, criterion (108) proposed in [8] cannot produce such a sharp synchronization condition.

For an analytical comparison between criterion (108), proposed in [8], and the new criterion (90)-(91), derived in this paper, a reasonable criterion is derived from (108) with $\varepsilon = 0$, as follows:

$$\begin{cases} k_{11} > -a, \\ k_{22} > \dfrac{(p_1 a + p_2 c)^2 - 2(p_1 a + p_2 c) p_2 \underline{\rho}_3 + p_2^2 \rho_3^2}{4 p_1 p_2 (k_{11} + a)}, \\ k_{33} > \dfrac{\rho_2^2 p_2^2 (k_{22} + 1)}{\Delta_2} - b, \end{cases} \tag{113}$$

where

$$\Delta_2 = 4 p_1 p_2 (k_{11} + a)(k_{22} + 1) - (p_1 a + p_2 c)^2 + 2(p_1 a + p_2 c) p_2 \underline{\rho}_3 - p_2^2 \rho_3^2, \tag{114}$$

and $\rho_2, \underline{\rho}_3$ and $\rho_3$ are defined as before. Only unidirectional coupling is considered here.

For the master Lorenz system defined by (111), one can obtain two synchronization conditions based on criterion (113) and criterion (90)-(91), respectively, *i.e.*,

$$\begin{cases} k_{11} = 1, \\ k_{22} > k_{22}^{'}(1) = 120.4, \\ k_{33} > k_{33}^{'}(1) = \dfrac{1406.25(k_{22} + 1)}{17(k_{22} + 1) - 2046.8} - 4, \end{cases} \tag{115}$$

which is related to criterion (113), and $p_1 = p_2 = 1$,

$$\begin{cases} k_{11} = 1, \\ k_{22} > k_{22}^{*}(1) = 54.6, \\ k_{33} > k_{33}^{*}(1) = \dfrac{77.65(k_{22} + 1)}{17(k_{22} + 1) - 944.16} - 4, \end{cases} \tag{116}$$

which is related to criterion (90)-(91).

For the master Lorenz system illustrated in Fig. 1, one has

$$\rho_2 = 24,\ \underline{\rho}_3 = 6.5,\ \rho_3 = \overline{\rho}_3 = 44.5\,. \tag{117}$$

Then, using criterion (113) with $p_1 = p_2 = 1$, one obtains

$$\begin{cases} k_{11} = 1, \\ k_{22} > k_{22}^{'}(2) = 65.6, \\ k_{33} > k_{33}^{'}(2) = \dfrac{144(k_{22}+1)}{11(k_{22}+1) - 732.6} - \dfrac{8}{3}. \end{cases} \tag{118}$$

Using criterion (90)-(91), one has

$$\begin{cases} k_{11} = 1, \\ k_{22} > k_{22}^{*}(2) = 31.0, \\ k_{33} > k_{33}^{*}(2) = \dfrac{29.4(k_{22}+1)}{11(k_{22}+1) - 352.5} - \dfrac{8}{3}. \end{cases} \tag{119}$$

It is clear that

$$\begin{cases} k_{22}^{*}(i) << k_{22}^{'}(i), \\ k_{33}^{*}(i) < k_{33}^{'}(i)\ for\ k_{22}(i) > k_{22}^{'}(i), \end{cases} \quad i = 1,2.$$

This means that the new criterion (90)-(91) is sharper than criterion (113), so is sharper than criterion (108), suggested by [8] and [9] respectively.

Very recently, a synchronization criterion for the scheme with bidirectional coupling is presented in Theorem 2 of [7], which is less conservative than the one proposed in Theorem 1 of [6].

If the unidirectional coupling with $K = diag\{k_{11}, k_{22}, k_{33}\}$ is considered, then their criterion can be reformulated as

$$\begin{cases} k_{11} > 0, k_{22} > 0, k_{33} > 0, \\ (k_{11}+a)(k_{22}+1)(k_{33}+b) > \dfrac{b^2(a+c)^2}{16(b-1)} N, \end{cases} \tag{120}$$

where

$$N = \max\{k_{22} + 1, k_{33} + b\}. \qquad (121)$$

Using our new criterion (90)-(91), one has

$$(k_{11} + a)(k_{22} + 1)(k_{33} + b) > \left[ \frac{a\rho_2^2}{4M} + \frac{a}{2}(M + c - \underline{\rho}_3) \right] N, \qquad (122)$$

where $N$ is defined by (121) and

$$M = \sqrt{\rho_3^2 - \underline{\rho}_3^2 + (c - \underline{\rho}_3)^2}.$$

To compare criterion (120) with criterion (122), choosing the Lorenz system illustrated in Fig. 1, one obtains

$$(k_{11} + 10)(k_{22} + 1)(k_{33} + 28) > 385.1N \qquad (123)$$

based on (120), and

$$(k_{11} + 10)(k_{22} + 1)(k_{33} + 28) > 357.4N \qquad (124)$$

based on (122).

This shows that the new criterion (122) is sharper than criterion (120), suggested in [7].

Take the coupling matrix $K = diag\{9,2,2\}$, which satisfies condition (124) but not (123). Figure 3 shows the performance of chaos synchronization of the master-slave Lorenz systems coupled by this matrix.

Fig.3

In [6] and [7], a synchronization scheme for two bidirectionally coupled classical Lorenz systems is considered, with $K_1 = diag\{d_{11}, 0, 0\}$ and $K_2 = diag\{d_{21}, 0, 0\}$, where $d_{11} = d_{21} = d_1$. A synchronization criterion is obtained as follows:

$$d_1 > \frac{b^2(a+c)^2}{32(b-1)} - \frac{a}{2}, \quad a \geq 1, \; b \geq 2. \tag{125}$$

If unidirectional coupling ($d_{11} = k_{11}, \; d_{21} = 0$) is used in the scheme, then the synchronization criterion is

$$k_{11} > \frac{b^2(a+c)^2}{16(b-1)} - a, \quad a \geq 1, \; b \geq 2. \tag{126}$$

Now, continue to use the system illustrated in Fig. 1 as example for simulation. Applying criterion (126) gives

$$k_{11} > k_{11}^{'}(3) = 375.0667\,. \tag{127}$$

From the new criterion (92), it follows that

$$k_{11} > k_{11}^{*}(3) = 353.245. \tag{128}$$

Thus, criterion (92) is sharper than criterion (126), proposed in [6] and [7], because $k_{11}^{*}(3) < k_{11}^{'}(3)$.

With the coupling matrix $K = diag\{354, 0, 0\}$, which satisfies (128) but not (127), Figure 4 illustrates the performance of chaos synchronization.

Fig.4

For the Lorenz system illustrated in Fig. 1, some synchronization criteria for the dislocated coupling matrix (72) and (84) were studied in [10].

However, there exist two errors on pages $203-204$ of [10]. The coefficients of the term $|e_1||e_2|$, $(10/\beta + 28 - k_2 + M_3)$ and $((10-k_1)/\beta + 28 + M_3)$ should be corrected as $(|10/\beta + 28 - k_2| + M_3)$ and $(|(10-k_1)/\beta + 28| + M_3)$, respectively. With the corrections, replacing $M_2, M_3, 1/\beta, k_1$ and $k_2$ by $\rho_2, \rho_3, \; p, k_{12}$ and $k_{21}$, respectively, one can revise the synchronization criteria described in Theorem 2 and Theorem 1 of [10],

as follows:

$$10+\frac{1}{p}(28+\rho_3)-\frac{1}{p}\sqrt{40p-\frac{3}{8}\rho_2^2} < k_{12} < 10\frac{1}{p}(28-\rho_3)+\frac{1}{p}\sqrt{40p-\frac{3}{8}\rho_2^2} \qquad (128)$$

with

$$p > p^{'}(1) = \frac{3\rho_2^2 + 8(\rho_3+28)^2}{320}, \qquad (129)$$

and

$$28+10p+\rho_3-\sqrt{40p-\frac{3}{8}\rho_2^2} < k_{12} < 28+10p-\rho_3+\sqrt{40p-\frac{3}{8}\rho_2^2} \qquad (130)$$

with

$$p > p^{'}(2) = \frac{3\rho_2^2 + 8\rho_3^2}{320}. \qquad (131)$$

Applying our new criteria (96) and (100), one can obtain the synchronization conditions for the dislocated coupling matrices as follows:

$$10+\frac{1}{p}(28-\underline{\rho}_3)-\frac{1}{p}\sqrt{40p-\frac{3}{8}\rho_2^2+\underline{\rho}_3^2-\rho_3^2} < k_{12} < 10+\frac{1}{p}(28-\overline{\rho}_3)+\frac{1}{p}\sqrt{40p-\frac{3}{8}\rho_2^2+\overline{\rho}_3^2-\rho_3^2} \qquad (132)$$

with

$$p > p^{*}(1) = \frac{3\rho_2^2 + 8\rho_3^2}{320}, \qquad (133)$$

and

$$28+10p-\underline{\rho}_3-\sqrt{40p-\frac{3}{8}\rho_2^2+\underline{\rho}_3^2-\rho_3^2} < k_{21} < 28+10p-\overline{\rho}_3+\sqrt{40p-\frac{3}{8}\rho_2^2+\overline{\rho}_3^2-\rho_3^2} \qquad (134)$$

with

$$p > p^{*}(2) = \frac{3\rho_2^2 + 8\rho_3^2}{320}. \qquad (135)$$

It is obvious that $p^{'}(1) > p^{*}(1) = p^{*}(2) = p^{'}(2)$. Thus, the new criterion (132) is sharper than criterion (128) because the synchronization area of the former covers that of the latter for any $p$ satisfying the constraint (129).

To compare criterion (130) with (134), first observe that

$$\begin{aligned}\sqrt{40p-\frac{3}{8}\rho_2^2+\bar{\rho}_3^2-\rho_3^2}-\bar{\rho}_3 &= \frac{40p-\frac{3}{8}\rho_2^2-\rho_3^2}{\sqrt{40p-\frac{3}{8}\rho_2^2+\bar{\rho}_3^2-\rho_3^2}+\bar{\rho}_3} \\ &\geq \frac{40p-\frac{3}{8}\rho_2^2-\rho_3^2}{\sqrt{40p-\frac{3}{8}\rho_2^2}+\rho_3} = \sqrt{40p-\frac{3}{8}\rho_2^2}-\rho_3 \end{aligned} \tag{136}$$

and that

$$\begin{aligned} -\underline{\rho}_3-\sqrt{40p-\frac{3}{8}\rho_2^2+\underline{\rho}_3^2-\rho_3^2} &< -\underline{\rho}_3-\sqrt{40p-\frac{3}{8}\rho_2^2}+\sqrt{\rho_3^2-\underline{\rho}_3^2} \\ &= \frac{\rho_3^2}{\sqrt{\rho_3^2-\underline{\rho}_3^2}+\underline{\rho}_3}-\sqrt{40p-\frac{3}{8}\rho_2^2} \\ &< \frac{\rho_3^2}{\rho_3+\underline{\rho}_3-|\underline{\rho}_3|}-\sqrt{40p-\frac{3}{8}\rho_2^2} \\ &\leq \rho_3-\sqrt{40p-\frac{3}{8}\rho_2^2}, \end{aligned} \tag{137}$$

where the following known inequality has been used:

$$\sqrt{a^2-b^2} > a-b, \quad \text{for} \quad a > b > 0.$$

Thus, the area determined by (134) covers that determined by (130). This means that our new criterion (134) is sharper than criterion (130).

Take $p = 138$, which satisfies (129) and (133), and $k_{12}$=9.8, which satisfies (132) but not (128) for the selected $p$. Figure 5 illustrates the performance of chaos synchronization for the dislocated coupling case.

Take $p = 55$, which satisfies (131) and (135), and $k_{21} = 566$, which satisfies (134)

but not (130) for $p = 55$. Figure 6 illustrates the performance of chaos synchronization for $k_{21} = 566$.

Fig.5 Fig.6

### 4.2. The Chen system

For the Chen system defined by (3) with parameter $a_{22} = c > 0$, Theorems 6, 10, 11, and 12 are not applicable.

Nevertheless, Theorem 4 and Theorem 8 lead to the following proposition.

**Proposition 2.** The master-slave synchronization scheme (6) for the Chen system achieves uniform chaos synchronization if either of the following synchronization criteria is satisfied:

(i) $K = diag\{k_{11}, k_{22}, k_{33}\}$ with

$$\begin{cases} k_{11} + a > 0, & (138) \\ k_{22} > c + \dfrac{\bar{\Delta}_c}{k_{11} + a}, & (139) \\ k_{33} > \dfrac{\rho_2^2 (k_{22} - c)}{4p_c^*((k_{11} + a)(k_{22} - c) - \bar{\Delta}_c)} - b, & (140) \end{cases}$$

where the parameters

$$[\bar{\Delta}_c, p_c^*] = \begin{cases} \left[\dfrac{a}{2}(\underline{u}_1 + c - a - \underline{\rho}_3),\ \dfrac{1}{a}\underline{u}_1\right], & if\ c - a \geq 0, or\ c - a < 0\ and \dfrac{\rho_3^2}{2(a-c)} \geq -\underline{\rho}_3, \\ \left[\dfrac{a}{2}(\bar{u}_1 + c - a - \bar{\rho}_3),\ \dfrac{1}{a}\bar{u}_1\right], & if\ c - a < 0\ and\ \dfrac{\rho_3^2}{2(a-c)} \geq -\bar{\rho}_3, \\ \left[\dfrac{\rho_3^2}{2(a-c)},\ 1 - \dfrac{c}{a}\right], & if\ c - a < 0\ and\ -\bar{\rho}_3 < \dfrac{\rho_3^2}{2(a-c)} < -\underline{\rho}_3, \end{cases} \tag{141}$$

with

$$\underline{u}_1 = \sqrt{\rho_3^2 - \underline{\rho}_3^2 + (\underline{\rho}_3 + a - c)^2}, \quad \bar{u}_1 = \sqrt{\rho_3^2 - \bar{\rho}_3^2 + (\bar{\rho}_3 + a - c)^2}; \tag{142}$$

(ii) $K = diag\{0, k_{22}, 0\}$ with

$$k_{22} > k_{22}^{*}(4) = c + L_m, \tag{143}$$

where,

$$L_m = \begin{cases} \dfrac{1}{2}\left(\underline{u}_2 + \dfrac{\rho_2^2}{4b} + c - a - \underline{\rho}_3\right), \ if \ \dfrac{\rho_2^2}{4b} + c - a \geq -\underline{u}_2, \\ \dfrac{1}{2}\left(\bar{u}_2 + \dfrac{\rho_2^2}{4b} - \dfrac{\rho_3^2 - \bar{\rho}_3^2}{a}\right), \ if \ \dfrac{\rho_2^2}{4b} + c - a \leq -\bar{u}_2, \\ \dfrac{ab\rho_3^2}{4ab\,|\,a - c\,| - \rho_2^2}, \ if \ -\bar{u}_2 < \dfrac{\rho_2^2}{4b} + c - a < -\underline{u}_2, \end{cases} \tag{144}$$

with

$$\underline{u}_2 = \sqrt{\left(\frac{\rho_2^2}{4b} + c - a - \underline{\rho}_3\right)^2 + \rho_3^2 - \underline{\rho}_3^2}, \quad \bar{u}_2 = \sqrt{\left(\frac{\rho_2^2}{4b} + c - a - \bar{\rho}_3\right)^2 + \rho_3^2 - \bar{\rho}_3^2}. \tag{145}$$

To verify these criteria, choose the chaotic Chen system with $a = 35, b = 3$ and $c = 28$, yielding a chaotic attractor of the system as shown in Fig. 7.

Fig.7

The system has the following bounds:

$$\rho_2 = 27, \ \underline{\rho}_3 = 6, \ \rho_3 = \bar{\rho}_3 = 44, \tag{146}$$

and satisfies $c{-}a{<}0$ and $\dfrac{\rho_3^2}{2(a-c)} \geq -\underline{\rho}_3$.

According to the new criterion consisting of (138)-(142), one can obtain the synchronization condition for $K = diag\{k_{11}, k_{22}, k_{33}\}$ as follows:

$$\begin{cases} k_{11} + 35 \geq 0, \\ k_{22} > 28 + \dfrac{568.5}{k_{11} + 35}, \\ k_{33} > \dfrac{140.2(k_{22} - 28)}{(k_{11} + 35)(k_{22} - 28) - 568.5} - 3. \end{cases} \tag{147}$$

With $K = diag\{5, 44, 28.5\}$, satisfying (147), Figure 8 illustrates the performance of chaos synchronization for the coupling matrix.

Fig.8

In [11], an optimized synchronization criterion for the master-slave Chen systems linearly coupled by $K = diag\{0, k_{22}, 0\}$ is given as follows:

$$k_{22} > k_{22}^{'}(4) = 2c + \frac{\rho_2^2}{4b} - a + \rho_3. \tag{148}$$

If the master Chen system satisfies

$$\frac{\rho_2^2}{4b} + c - a \geq 0,$$

then the critical value of the criterion defined by (143)-(145) is

$$\begin{aligned} k_{22}^{*}(4) &= c + \frac{1}{2}\left(\sqrt{\left(\frac{\rho_2^2}{4b} + c - a - \underline{\rho}_3\right)^2 + \rho_3^2 - \underline{\rho}_3^2} + \frac{\rho_2^2}{4b} + c - a - \underline{\rho}_3\right) \\ &< c + \frac{1}{2}\left(\left|\frac{\rho_2^2}{4b} + c - a - \underline{\rho}_3\right| + \sqrt{\rho_3^2 - \underline{\rho}_3^2} + \frac{\rho_2^2}{4b} + c - a - \underline{\rho}_3\right) \\ &< c + \frac{\rho_2^2}{4b} + c - a + \frac{1}{2}\left(\left|\underline{\rho}_3\right| - \underline{\rho}_3 + \sqrt{\rho_3^2 - \underline{\rho}_3^2}\right) < k_{22}^{'}(4). \end{aligned}$$

This means that the new criterion (143)-(145) is sharper than criterion (148).

Using criterions (143)-(145) and (148), one can solve for the critical values for the Chen system, resulting in $k_{22}^{'}(4) = 125.8$ and $k_{22}^{*}(4) = 84.2$, respectively. With the coupling matrix $K = diag\{0, 85, 0\}$, which satisfies (143)-(145) but not (148), Figure 9 illustrates the performance of chaos synchronization.

Fig.9

### 4.3. The Lü system

Since $a_{22} = c > 0$ in the Lü system, only Theorem 4 and Theorem 8 can be utilized to deduce some synchronization criteria.

**Proposition 3.** The master-slave synchronization scheme (6) for the Lü system achieves uniform chaos synchronization if either of the following synchronization criteria is satisfied:

(i) $K = diag\{k_{11}, k_{22}, k_{33}\}$ with

$$\begin{cases} k_{11} + a > 0, \\ k_{22} > c + \dfrac{a(\rho_3 - \underline{\rho}_3)}{2(k_{11} + a)}, \\ k_{33} > \dfrac{a\rho_2^2 (k_{22} - c)}{4\rho_3 \overline{\Delta}_k} - b, \end{cases} \quad (149)$$

where

$$\overline{\Delta}_k = (k_{11} + a)(k_{22} - c) - \frac{a}{2}(\rho_3 - \underline{\rho}_3); \quad (150)$$

(ii) $K = diag\{0, k_{22}, 0\}$ with

$$k_{22} > k_{22}^*(5) = c + \frac{1}{2}\left(\sqrt{(\frac{\rho_2^2}{4b} - \underline{\rho}_3)^2 + \rho_3^2 - \underline{\rho}_3^2} + \frac{\rho_2^2}{4b} - \underline{\rho}_3\right). \quad (151)$$

Choosing the Lü system with $a = 36, b = 3$ and $c = 20$ gives a chaotic attractor as shown in Fig. 10.

Fig.10

One can obtain the bounds of this system from Fig. 10, as follows:

$$\rho_2 = 24.5, \underline{\rho}_3 = 4, \rho_3 = \overline{\rho}_3 = 39. \quad (152)$$

Applying the new criterion (149) to this system results in a synchronization condition for $K = diag\{k_{11}, k_{22}, k_{33}\}$, as

$$\begin{cases} k_{11}+36>0, \\ k_{22}>20+\dfrac{630}{k_{11}+36}, \\ k_{33}>\dfrac{138.5(k_{22}-20)}{(k_{11}+36)(k_{22}-20)-630}-3. \end{cases} \tag{153}$$

It is easy to verify that $K = diag\{-1, 40, 37\}$ satisfies condition (153). Figure 11 illustrates the performance of chaos synchronization for this case.

Fig.11

In [12] and [13], the following synchronization criterion for two Lü systems coupled by $K = diag\{0, k_{22}, 0\}$ is given:

$$k_{22} > k_{22}^{'}(5) = \frac{\rho_2^2}{4b} + c + \rho_3. \tag{154}$$

If the master Lü system has a bound $\underline{\rho}_3 \geq 0$, then the critical value of criterion (151) is

$$k_{22}^{*}(5) < c + \frac{1}{2}\left( \left| \frac{\rho_2^2}{4b} - \underline{\rho}_3 \right| + \sqrt{\rho_3^2 - \underline{\rho}_3^2} + \frac{\rho_2^2}{4b} - \underline{\rho}_3 \right) < c + \frac{\rho_2^2}{4b} + \frac{1}{2}\sqrt{\rho_3^2 - \underline{\rho}_3^2} < k_{22}^{'}(5).$$

If $\underline{\rho}_3 < 0$, then

$$k_{22}^{*}(5) < c + \frac{1}{2}\left( \frac{\rho_2^2}{2b} + 2|\underline{\rho}_3| + \sqrt{\rho_3^2 - \underline{\rho}_3^2} \right) = c + \frac{\rho_2^2}{4b} + |\underline{\rho}_3| + \frac{1}{2}\sqrt{\rho_3^2 - \underline{\rho}_3^2} < k_{22}^{'}(5).$$

Thus, the new criterion (151) is sharper than criterion (154).

For the Lü system illustrated in Fig.10, one can obtain the critical values as $k_{22}^{*}(5) = 73.1$ and $k_{22}^{'}(5) = 109.0$. With the coupling matrix $K = diag\{0, 74, 0\}$, which satisfies our criterion (151) but not (154), Figure 12 illustrates the performance of chaos synchronization.

Fig.12

### 4.4. A unified chaotic system

For the unified chaotic system defined by (5), one can derive the synchronization criteria based on Theorems 4, 6, 8, 10, 11, and 12.

**Proposition 4.** The master-slave synchronization scheme (6) for the unified chaotic system achieves uniform chaos synchronization if any of the following synchronization criteria is satisfied:

(i) $K = diag\{k_{11}, k_{22}, k_{33}\}$ with

$$\begin{cases} k_{11} > -(25\alpha+10), \\ k_{22} > 29\alpha - 1 + \dfrac{25\alpha+10}{2(k_{11}+25\alpha+10)}(\underline{u}_3 + 28 - 35\alpha - \underline{\rho}_3), \\ k_{33} > -\dfrac{\alpha+8}{3} + \dfrac{\rho_2^2(25\alpha+10)(k_{22}+1-29\alpha)}{4\underline{u}_3\bar{\Delta}_u} - b \ , \end{cases} \tag{155}$$

where $\underline{\rho}_3 \geq 0, \alpha \in [0,1]$, and

$$\bar{\Delta}_u = (k_{11}+25\alpha+10)(k_{22}+1-29\alpha) - \frac{25\alpha+10}{2}(\underline{u}_3 + 28 - 35\alpha - \underline{\rho}_3), \tag{156}$$

$$\underline{u}_3 = \sqrt{\rho_3^2 - \underline{\rho}_3^2 + (\underline{\rho}_3 + 35\alpha - 28)^2} \ ; \tag{157}$$

(ii) $K = diag\{k_{11}, 0, 0\}$ with

$$k_{11} > \frac{25\alpha+10}{2(1-29\alpha)}(\underline{u}_4 + 23\alpha + 26 - \underline{\rho}_3), \tag{158}$$

where $\alpha \in [0, 1/29)$ and

$$\underline{u}_4 = \sqrt{\rho_3^2 - \underline{\rho}_3^2 + (\underline{\rho}_3 + 35\alpha - 28)^2 + \frac{3(1-29\alpha)}{\alpha+8}\rho_2^2} \ ; \tag{159}$$

(iii) $K = diag\{0, k_{22}, 0\}$ with

$$k_{22} > 29\alpha - 1 + \frac{1}{2}\sqrt{(g(\alpha)-\underline{\rho}_3)^2 + \rho_3^2 - \underline{\rho}_3^2} + g(\alpha) - \underline{\rho}_3), \tag{160}$$

where

$$g(\alpha)=\frac{3\rho_2^2}{4(\alpha+8)}+28-35\alpha, \tag{161}$$

$$g(\alpha)+\sqrt{(g(\alpha)-\underline{\rho}_3)^2+\rho_3^2-\underline{\rho}_3^2}>0. \tag{162}$$

(iv) $K=diag\{0,0,k_{33}\}$ with

$$k_{33}>-\frac{\alpha+8}{3}+\frac{1-29\alpha}{Q_m}\rho_2^2, \tag{163}$$

where $\alpha\in[0,1/29)$, and

$$\underline{\rho}_3-23\alpha-26>\sqrt{(\underline{\rho}_3+35\alpha-28)^2+\rho_3^2-\underline{\rho}_3^2}, \tag{164}$$

$$Q_m=4(29\alpha-1)(\underline{\rho}_3+25-52\alpha)-\rho_3^2-\underline{+\rho}_3^2. \tag{165}$$

It should be noted that the above proposition only gives one part of synchronization criteria for the unified chaotic system. For the case of $\underline{\rho}_3<0$, or when inequality (162) is not satisfied, the synchronization criteria for $K=diag\{k_{11},k_{22},k_{33}\}$ and $K=diag\{0,k_{22},0\}$ can be derived from Theorem 4 and Theorem 8, respectively, and the criteria for the dislocated coupling matrix (72) and (84) can be derived from Theorem 11 and Theorem 12, respectively.

In [7] and [14], some synchronization criteria for the unified chaotic systems bidirectionally coupled by $K_1=K_2=diag\{d_1,d_2,d_3\}$ are derived. If the unidirectional coupling is used, criterion (9) of [7] can be re-stated as

$$\begin{cases} k_{11}>0,\ k_{22}>28,\ k_{33}>0, \\ (k_{11}+25\alpha+10)(k_{22}+1-29\alpha)\left(k_{33}+\dfrac{8+\alpha}{3}\right)>\left(\dfrac{M}{2}\right)^2(k_{22}+1-29\alpha)+\left(19-5\alpha+\dfrac{M}{2}\right)^2\left(k_{33}+\dfrac{8+\alpha}{3}\right)=k'(6), \end{cases} \tag{166}$$

and criterion (8) of [14] is equivalent to

$$\begin{cases} k_{11}+25\alpha+10>\dfrac{\sqrt{2}M}{2}, \\ k_{22}+1-29\alpha>\dfrac{(19-5\alpha)^2}{k_{11}+25\alpha+10-\dfrac{\sqrt{2}M}{2}}+\dfrac{\sqrt{2}M}{2}, \\ k_{33}+\dfrac{8+\alpha}{3}>\dfrac{\sqrt{2}M}{2}, \end{cases} \tag{167}$$

where $M=\max\{|x_i|, i=1,2,3\}=\max\{\rho_1,\rho_2,\rho_3\}$.

To compare the new criterion (155) with criterion (166), first produce the following criterion from (155):

$$\begin{cases} k_{11}>0,\ k_{22}>28,\ k_{33}>0, \\ (k_{11}+25\alpha+10)(k_{22}+1-29\alpha)\left(k_{33}+\dfrac{8+\alpha}{3}\right) \\ >\dfrac{\rho_2^2(25\alpha+10)}{4\underline{u}_3}(k_{22}+1-29\alpha)+\dfrac{25\alpha+10}{2}(\underline{u}_3+28-35\alpha-\underline{\rho}_3)\left(k_{33}+\dfrac{8+\alpha}{3}\right)=k^*(6). \end{cases} \tag{168}$$

Take $\alpha=0$, 1, and 0.8, which are corresponding to the classical Lorenz system, Chen system and Lü system, respectively. Their bounds are defined respectively by (117), (146) and (152). The critical values $k^*(6)$ and $k^{'}(6)$ are calculated and listed in Table 1.

Tab.1

It is apparent that the new criterion (168) is sharper than criterion (166) according to the comparison given in Table 1. In fact, the synchronization area determined by criterion (155) is broader than that by (168).

For $\alpha=0$ (the Lorenz system), take the coupling matrix $K=diag\{50,38,1\}$,

which satisfies our new criterion (168) but not criterion (166). The performance of chaos synchronization for this coupling is illustrated in Fig. 13.

Fig.13

In [14], the following optimization problem is considered:

$$\begin{cases} \min\limits_{k_{11},k_{22},k_{33}} k_{11}+k_{22}+k_{33}, \\ \text{subjected to (167).} \end{cases} \tag{169}$$

Using numerical methods, the optimized coupling matrices satisfying (169) are obtained as follows [14]:

(i) $K = diag\{14.3556,\ 77.3554,\ 32.3554\}$ for $\alpha$=1;

(ii) $K = diag\{41.3552,\ 55.7554,\ 32.6554\}$ for $\alpha$=0.1.

The bounds of the system with $\alpha = 1$ are determined by (146) and that with $\alpha = 0.1$ can be obtained from Fig. 14 as follows:

$$\rho_2 = 25.5,\ \underline{\rho}_3 = 6.0,\ \rho_3 = \bar{\rho}_3 = 46. \tag{170}$$

Fig.14

Using the new criterion (155), the following synchronization conditions can be obtained:

(i) $K = diag\{14.3556,\ 50,\ 3.7\}$, for $\alpha = 1$;

(ii) $K = diag\{30,\ 15,\ 1.5\}$, for $\alpha = 0.1$.

This shows that the new criterion (155) is sharper than criterion (167).

For the unified chaotic system with $\alpha$=0.1, take the coupling matrix $K = diag\{30,\ 15,\ 1.5\}$. Figure 15 illustrates the performance of chaos synchronization.

Fig.15

## 5. Conclusion

Chaos synchronization of the general master-slave identical generalized Lorenz systems via linear state error feedback control has been systematically investigated in this paper. By means of linearization and Lyapunov's direct methods, some new sufficient synchronization criteria have been obtained, which are more flexible and less conservative than those given in the literature. Optimization techniques are applied to these criteria so that the ranges of the coupling coefficients, determined by the criteria, are enlarged, improving the sharpness of the criteria. As application of the theoretical results, synchronization criteria for four typical generalized Lorenz systems have been derived in algebraic forms, which are easy to verify. It is believed that the main results of the paper provide a promising new approach for synchronization of autonomous chaotic systems via the simple but effective linear state error feedback control.

## Acknowledgements

Research is supported by National Nature Science Foundation of China under grant No. 60674049 and China Postdoctoral Science Foundation under grant No. 2005038177.

**Figure**

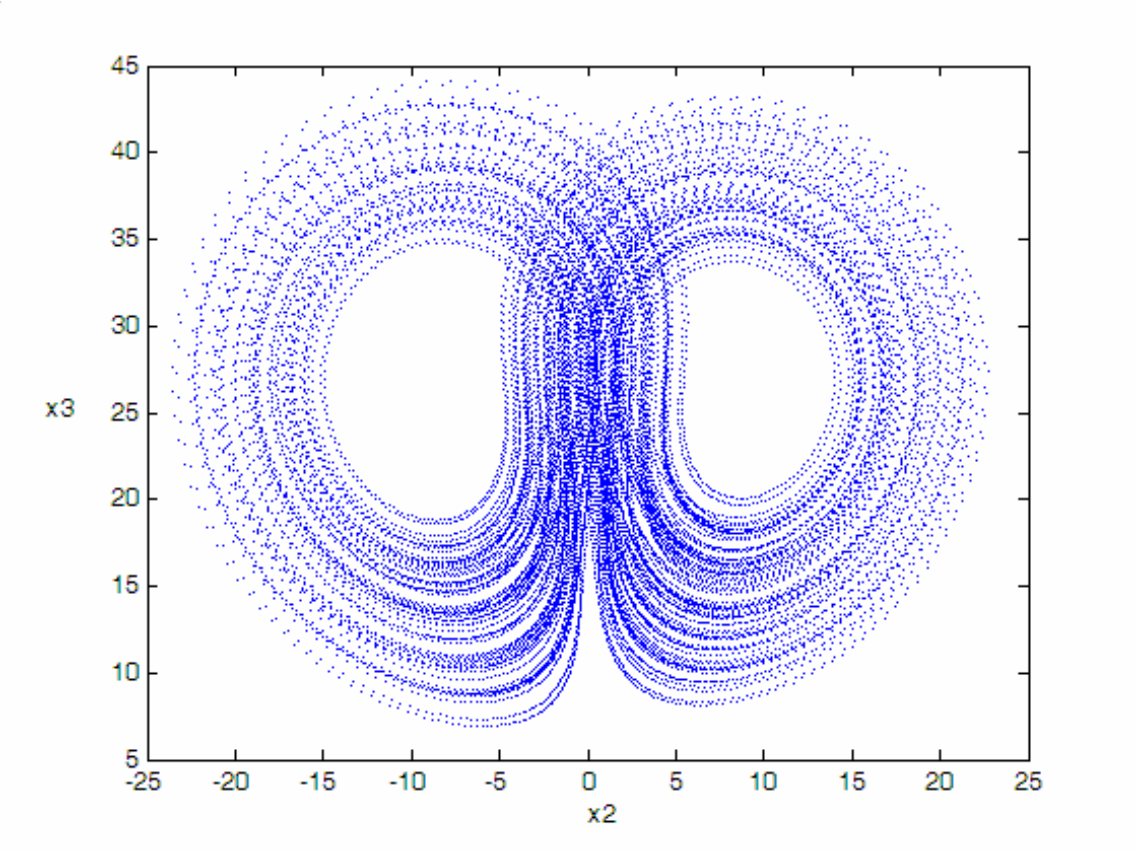

Fig. 1. The chaotic attractor of the classical Lorenz system in the $x_2 - x_3$ plane, with parameters $a = 10$, $b = 8/3$, $c = 28$, and initial condition $x(0) = (2, 3.5, 18.4)$.

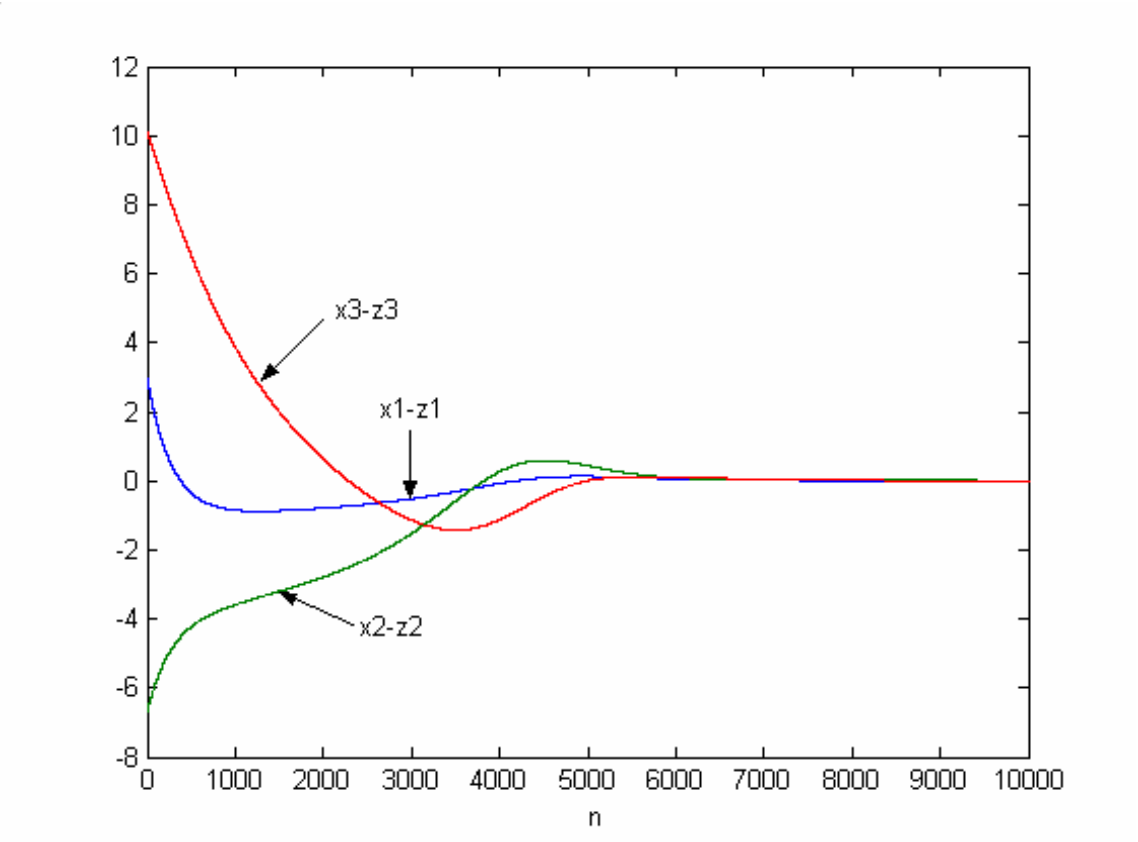

Fig.2. Chaos synchronization of two Lorenz systems coupled by $K = diag\{45.6, 16, 12.9\}$, with parameters $a = 16$, $b = 4$, $c = 45.6$, and initial conditions $x(0) =$ $(2, 3.5, 18.4)$, $z(0) = (-1, 10.2, 8.3)$.

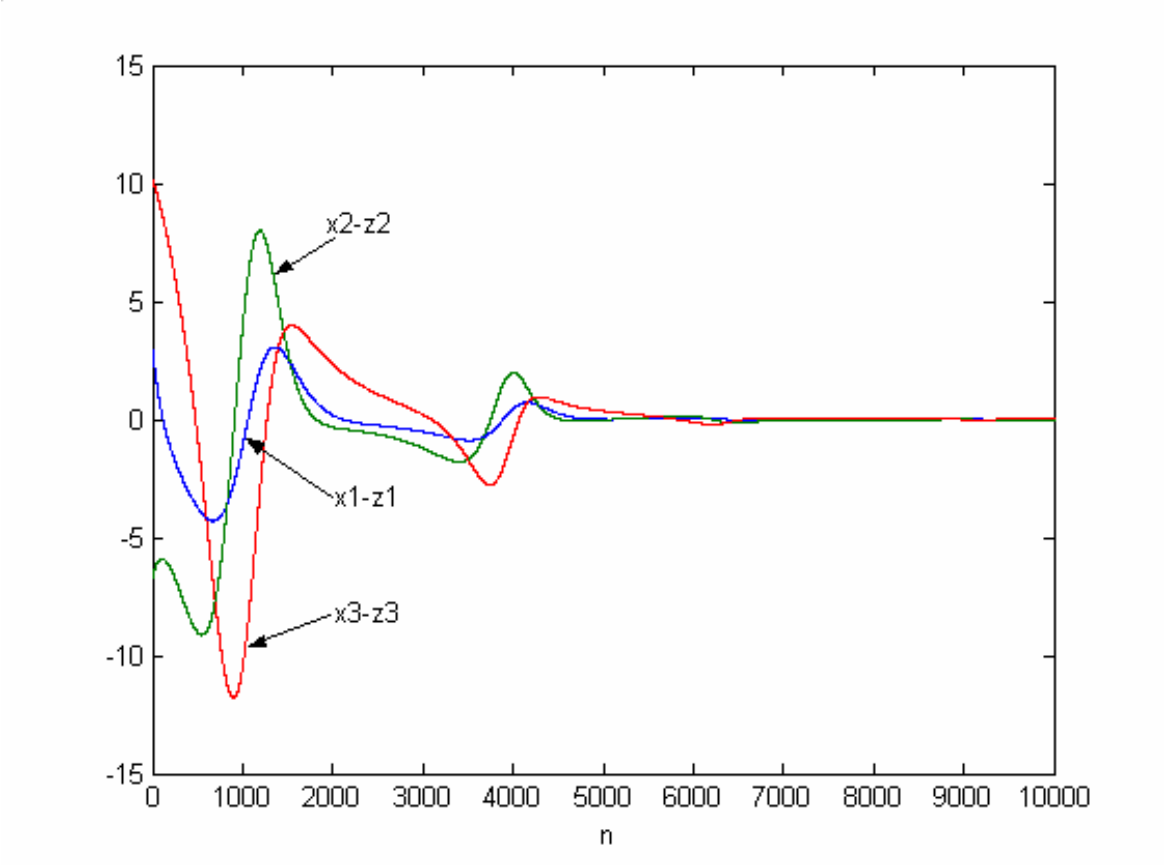

Fig. 3. Chaos synchronization of two Lorenz systems coupled by $K = diag\{9, 2, 2\}$, with parameters $a=10$, $b=8/3$, $c=28$, and initial conditions $x(0) = (2, 3.5, 18.4)$, $z(0) = (-1, 10.2, 8.3)$.

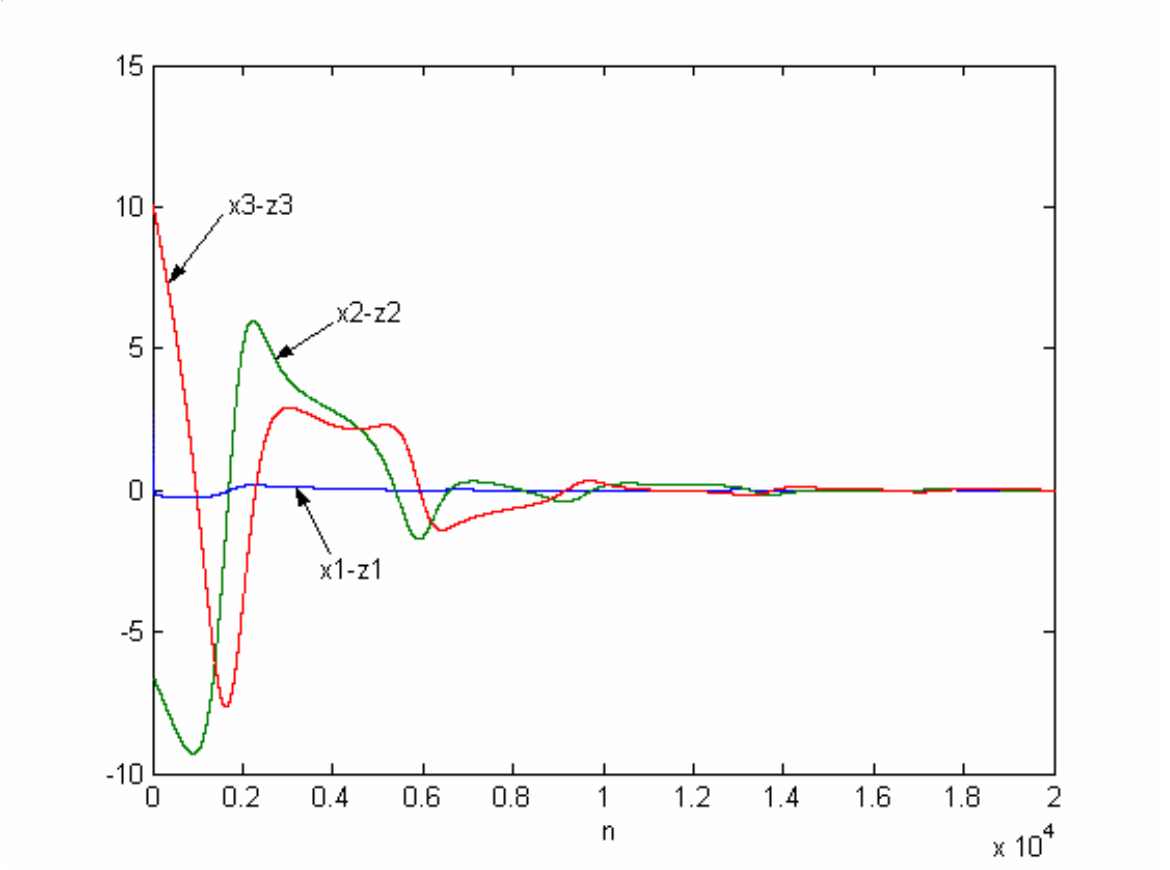

Fig. 4. Chaos synchronization of two Lorenz systems coupled by $K = diag\{354, 0, 0\}$, with parameters $a=10$, $b=8/3$, $c=28$, and initial conditions $x(0) = (2, 3.5, 18.4)$, $z(0) = (-1, 10.2, 8.3)$.

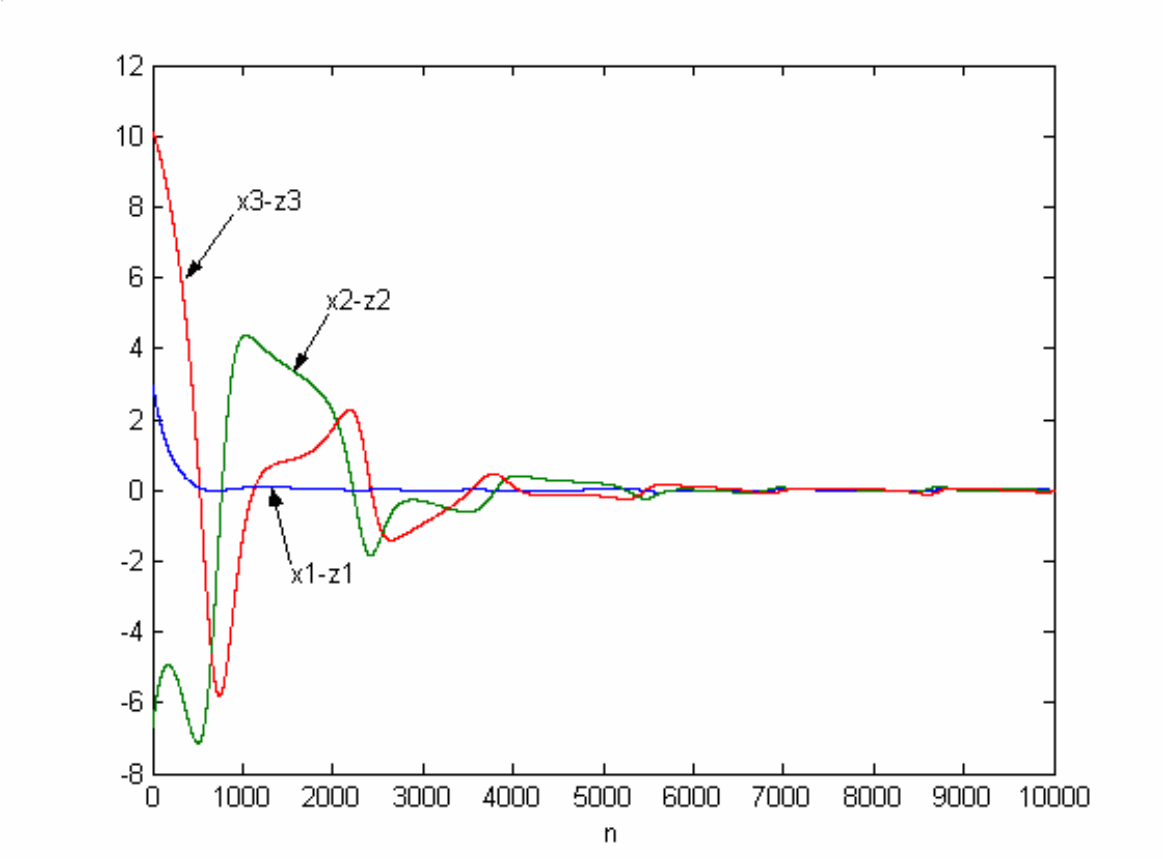

Fig. 5. Chaos synchronization of two Lorenz systems coupled by the coupling matrix (72) with $k_{12}=9.8$ , where $a=10$, $b=8/3$, $c=28$, and initial conditions $x(0)=(2, 3.5, 18.4)$, $z(0)=(-1, 10.2, 8.3)$.

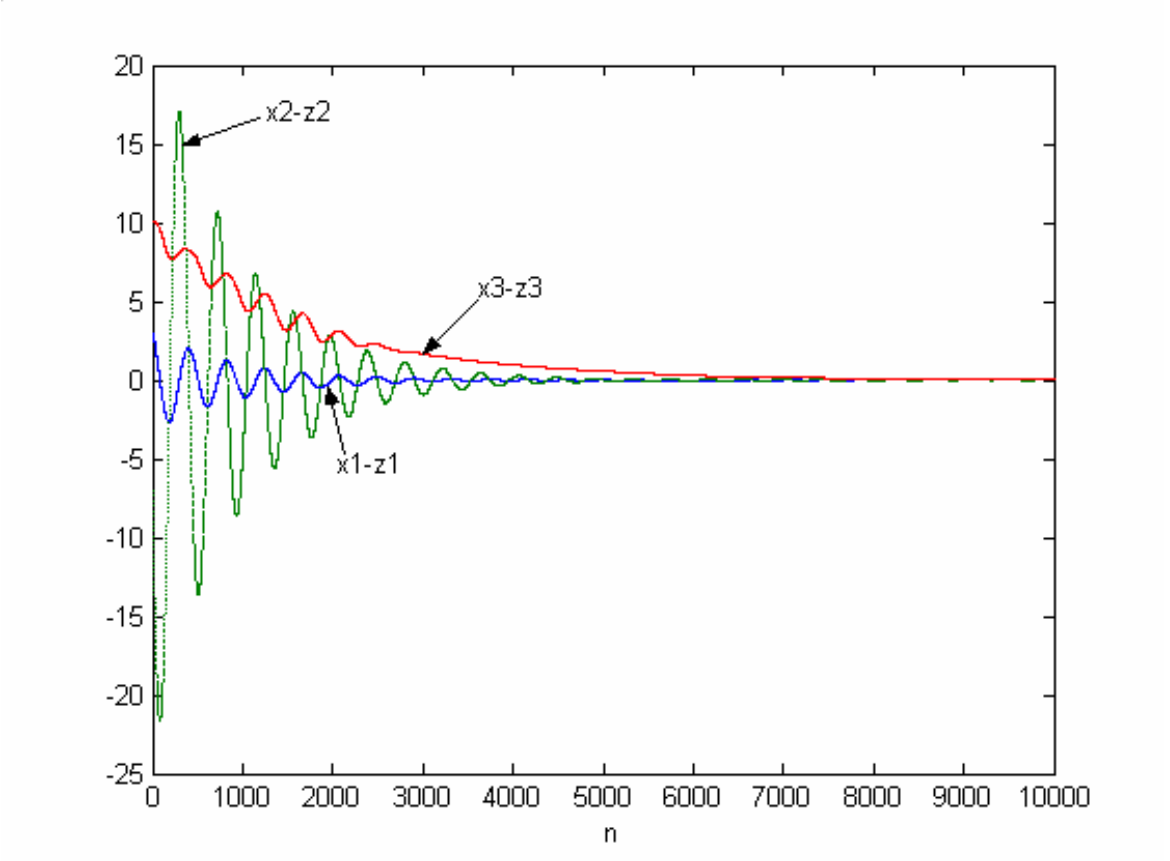

Fig. 6. Chaos synchronization of two Lorenz systems coupled by the coupling matrix (84) with $k_{21}=566$ , where $a=10$, $b=8/3$, $c=28$, and initial conditions $x(0)=(2, 3.5, 18.4)$, $z(0)=(-1, 10.2, 8.3)$.

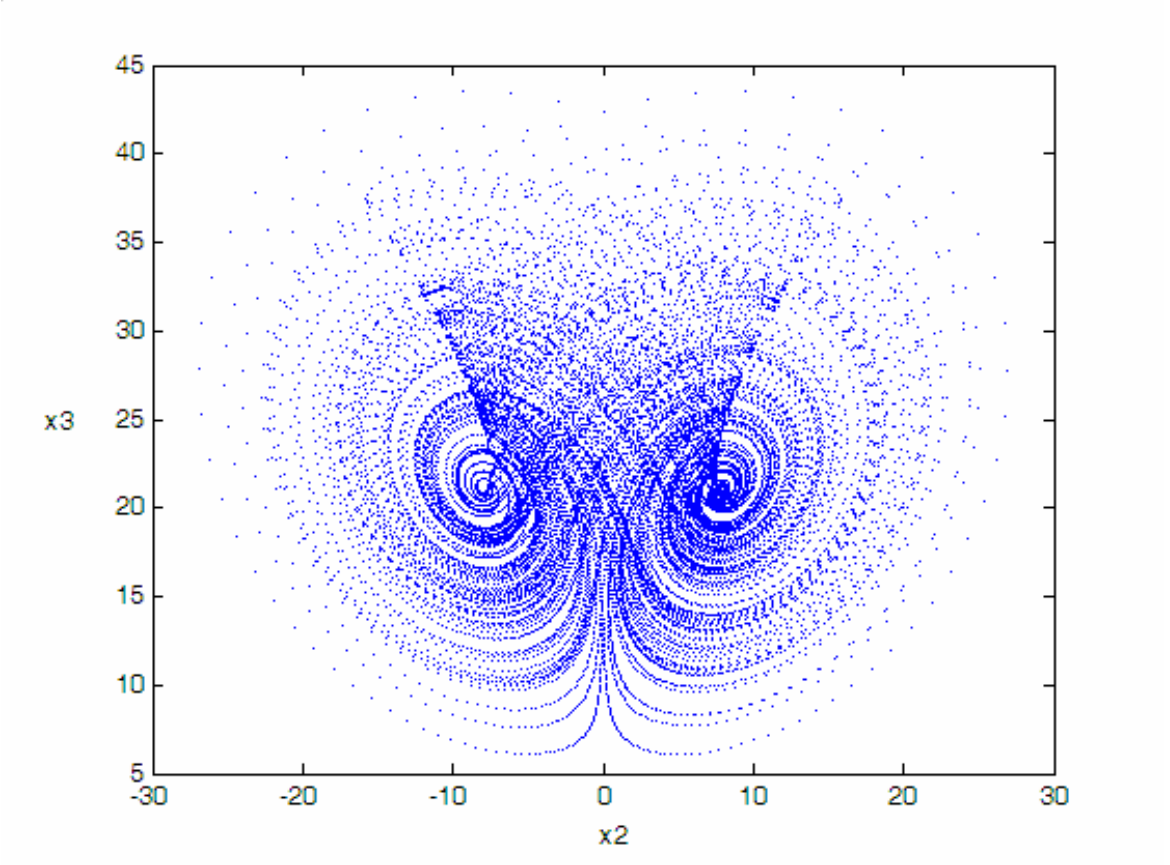

Fig. 7. The chaotic attractor of Chen system in the $x_2 - x_3$ plane, with parameters $a = 35$, $b = 3$, $c = 28$, and initial condition $x(0) = (3, -5.4, 35.6)$.

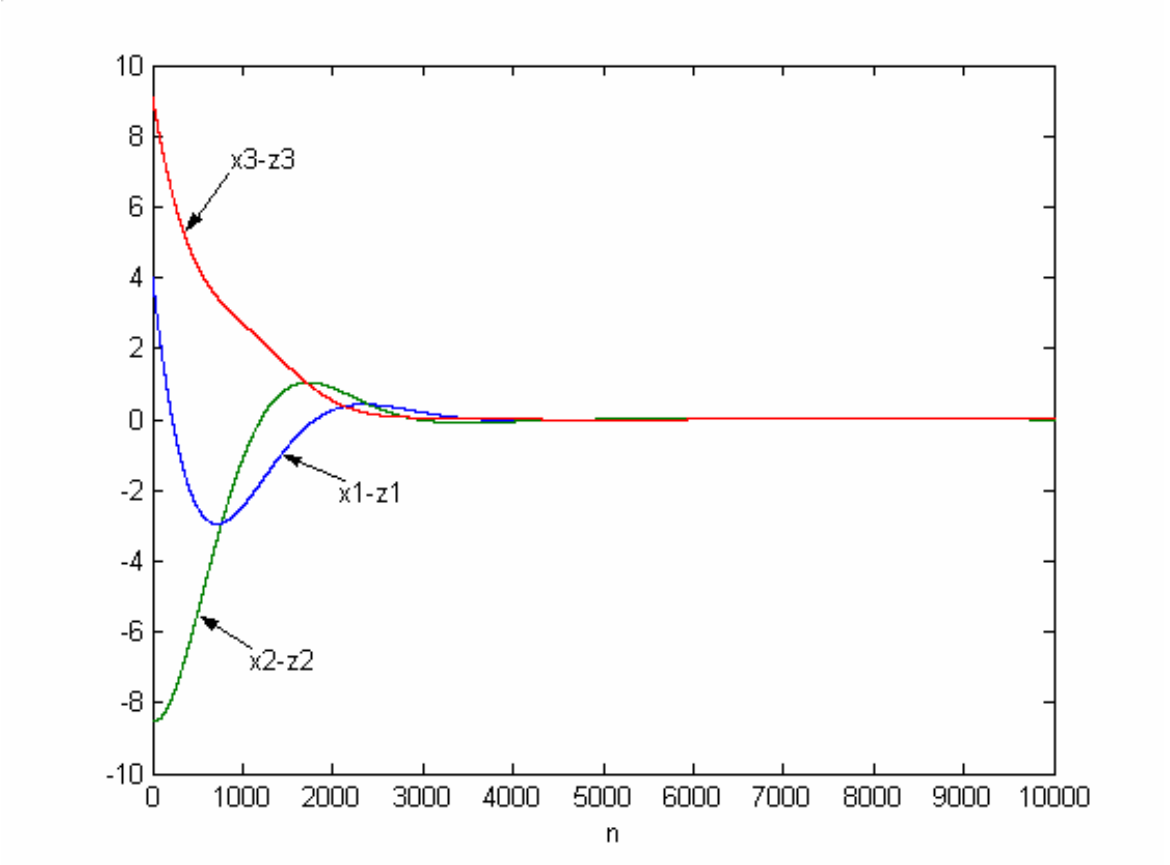

Fig. 8. Chaos synchronization of two Chen systems coupled by $K = diag\{5, 44, 28.5\}$, with parameters $a = 35$, $b = 3$, $c = 28$, and initial condition $x(0) = (3, -5.4, 35.6)$, $z(0) = (-1, 3.1, 26.5)$.

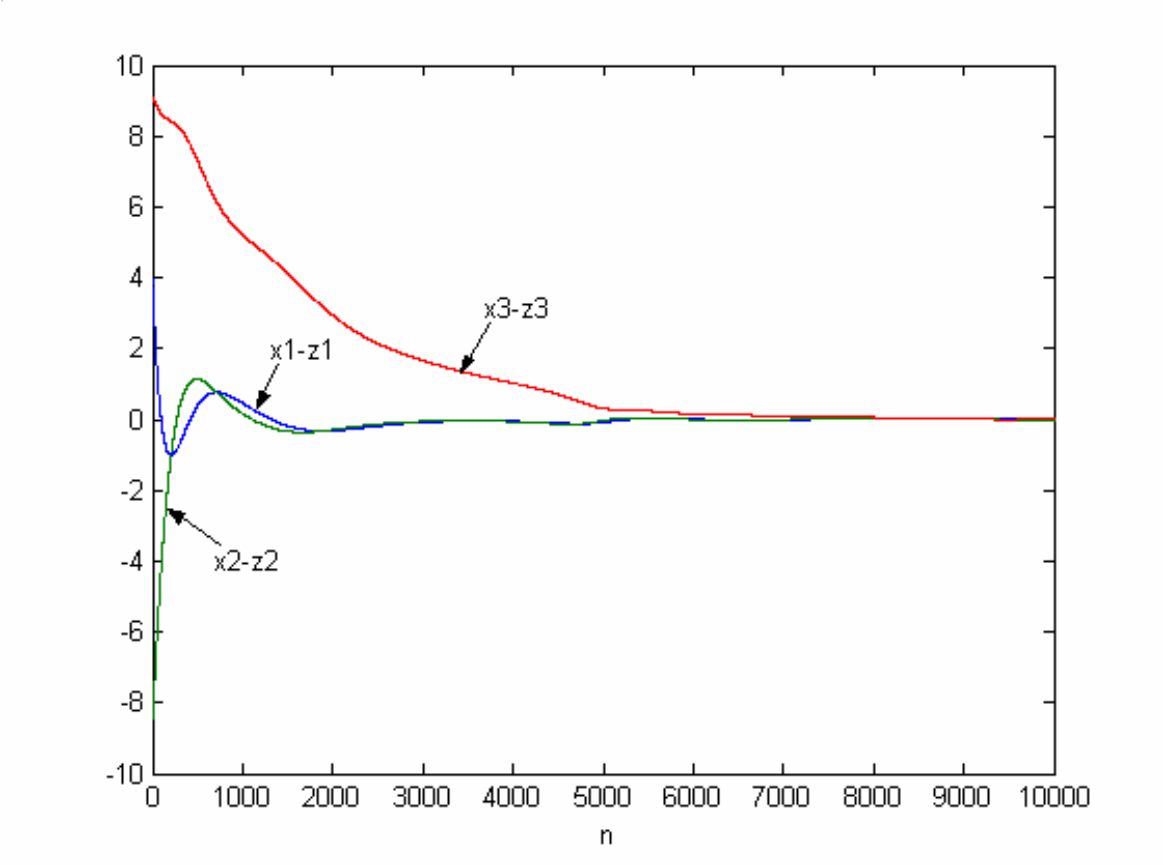

Fig. 9. Chaos synchronization of two Chen systems coupled by $K = diag\{0, 85, 0\}$, with parameters $a=35$, $b=3$, $c=28$, and initial condition $x(0) = (3, -5.4, 35.6)$, $z(0) = (-1, 3.1, 26.5)$.

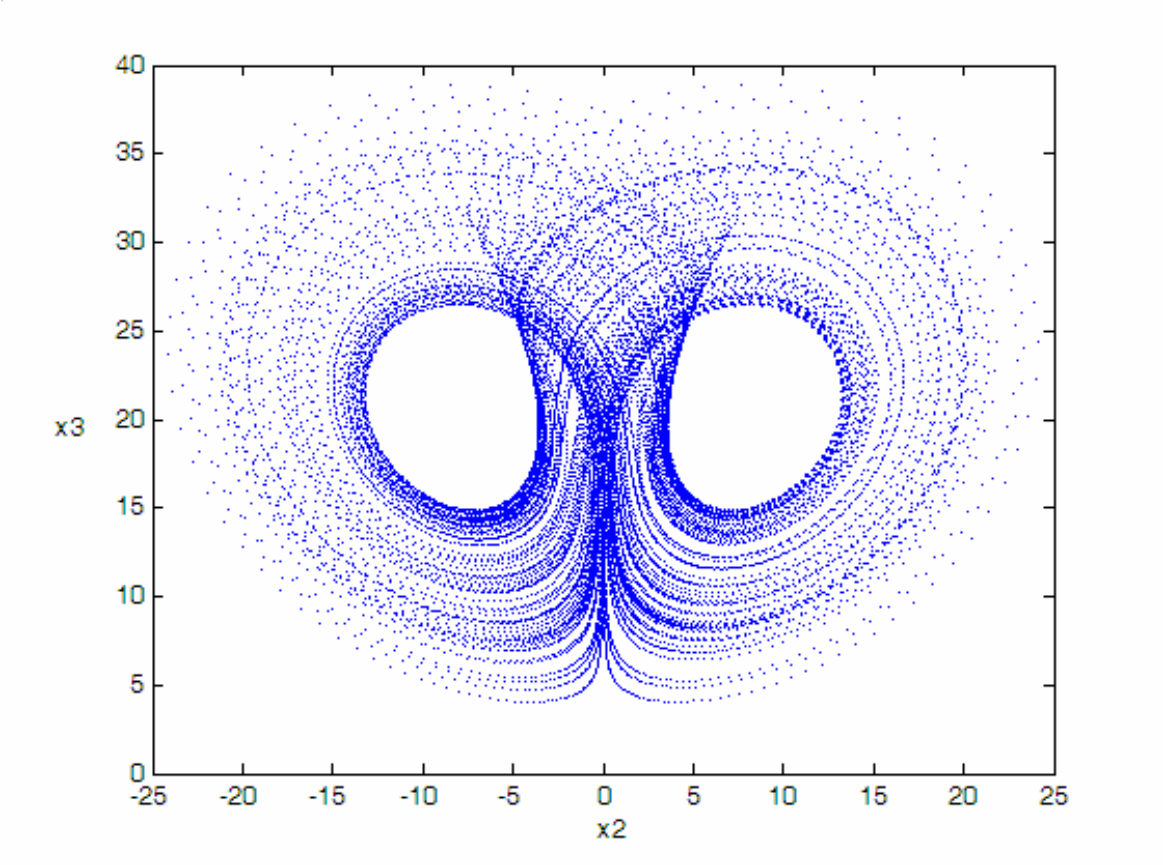

Fig. 10. The chaotic attractor of Lü system in the $x_2 - x_3$ plane, with parameters $a=36$, $b=3$, $c=20$, and initial condition $x(0) = (-4.2, -3.4, 20.3)$.

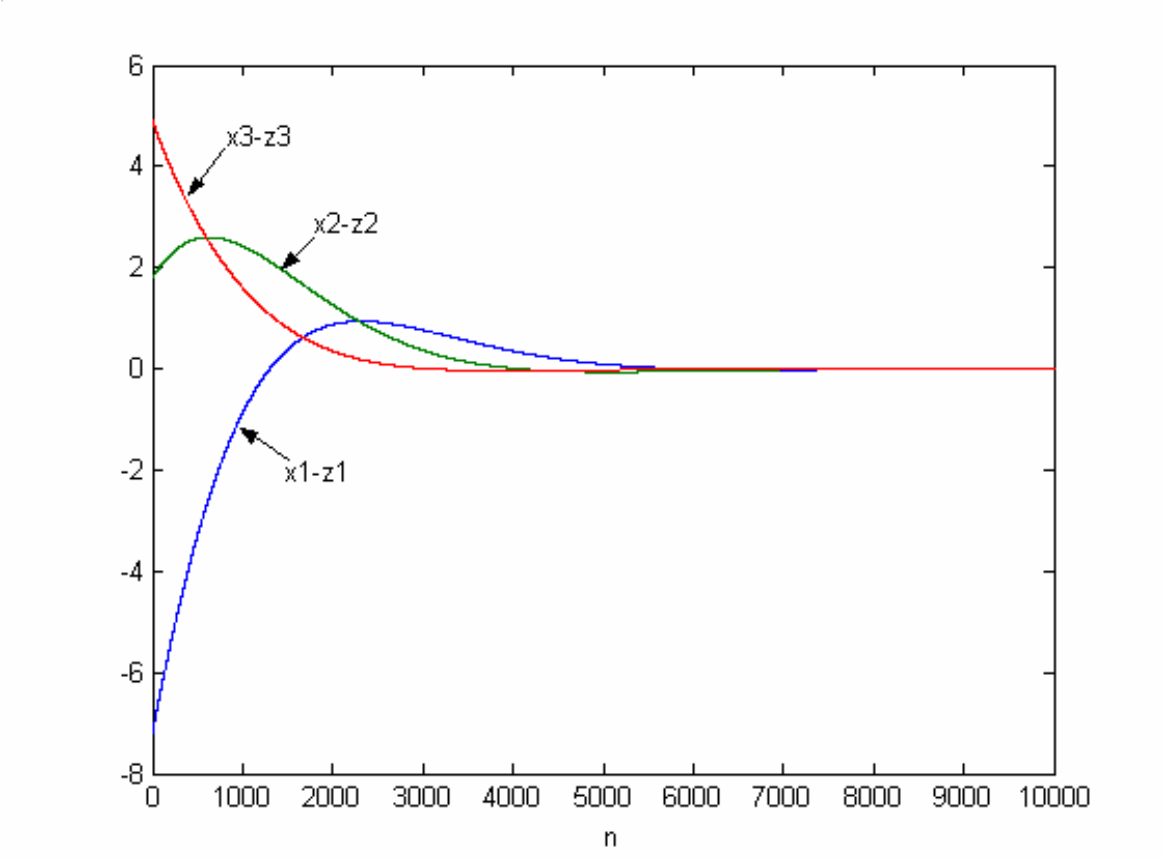

Fig. 11. Chaos synchronization of two Lü systems coupled by $K = diag\{-1, 40, 37\}$, with parameters $a$=36, $b$=3, $c$=20, and initial condition $x(0) = (-4.2, -3.4, 20.3)$, $z(0) = (3, -5.2, 15.4)$.

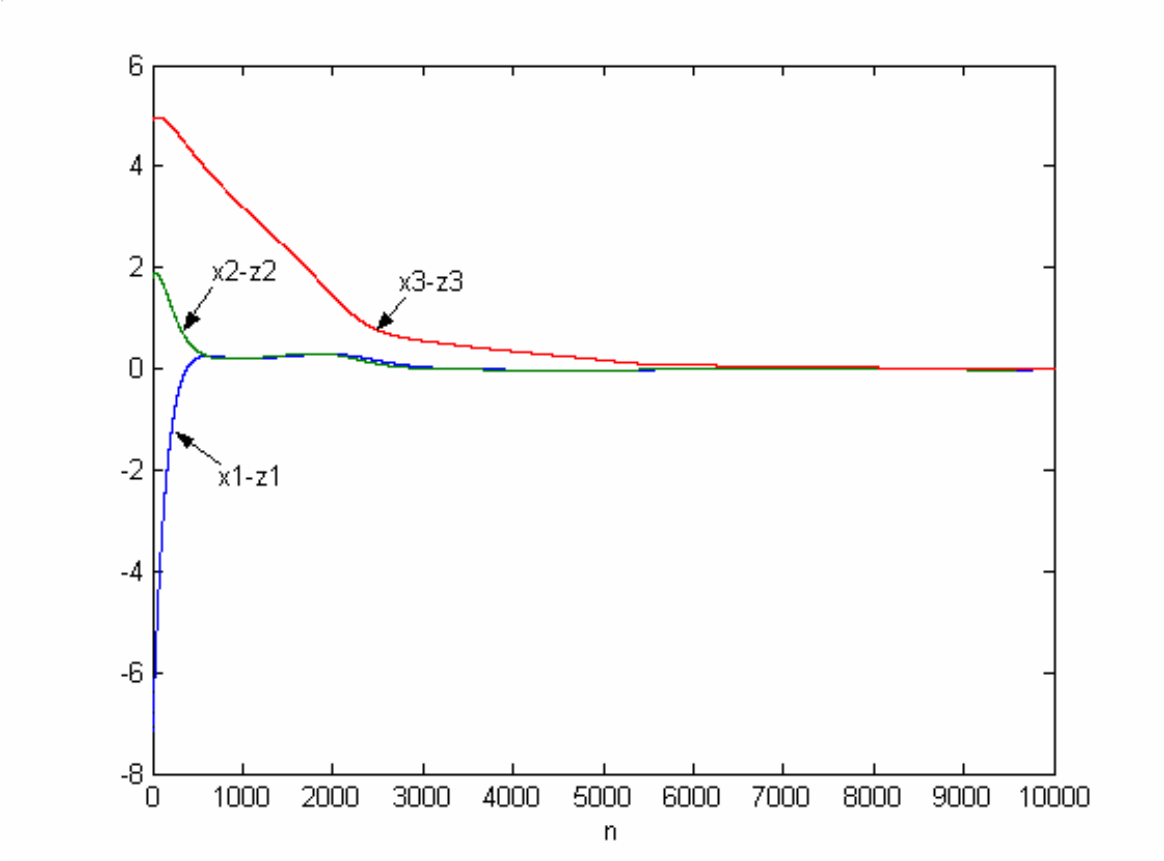

Fig. 12. Chaos synchronization of two Lü systems coupled by $K = diag\{0, 74, 0\}$, with parameters $a$=36, $b$=3, $c$=20, and initial condition $x(0) = (-4.2, -3.4, 20.3)$, $z(0) = (3, -5.2, 15.4)$.

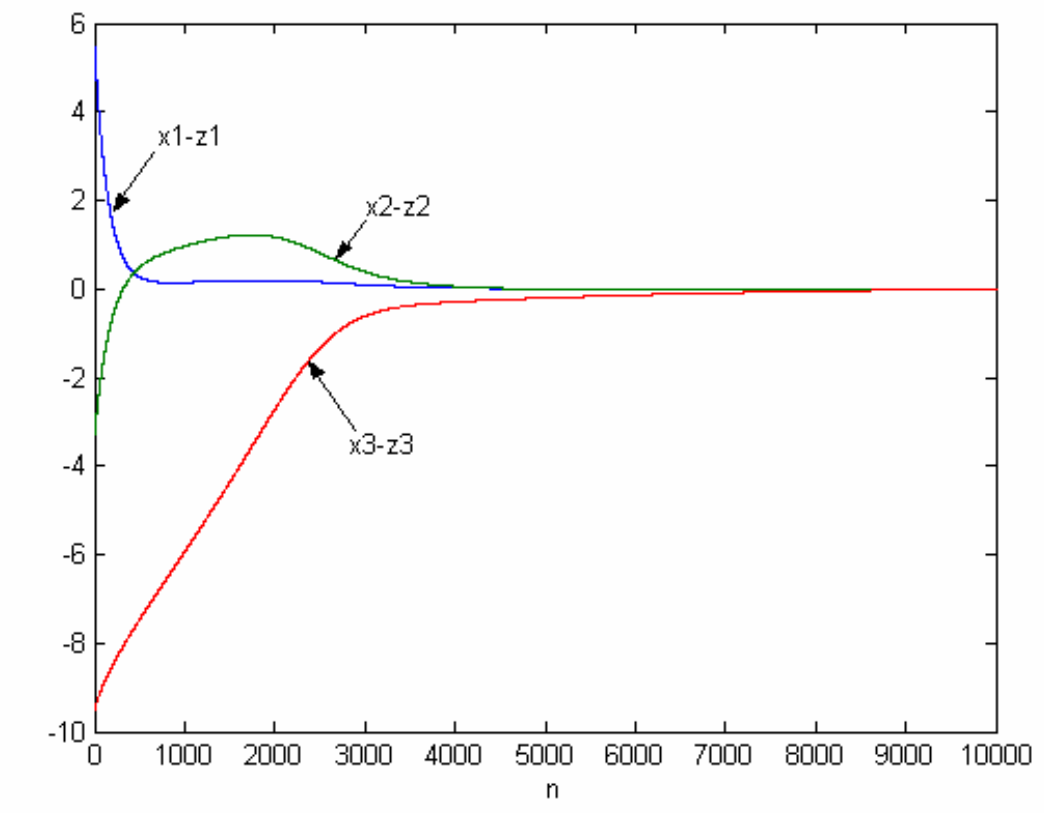

Fig. 13. Chaos synchronization of two unified chaotic systems coupled by $K = diag\{50, 38, 1\}$, with parameters $\alpha = 0$, and initial condition $x(0) = (3,\ 5,\ 7)$, $z(0) = (-2.5,\ 8.3,\ 16.5)$.

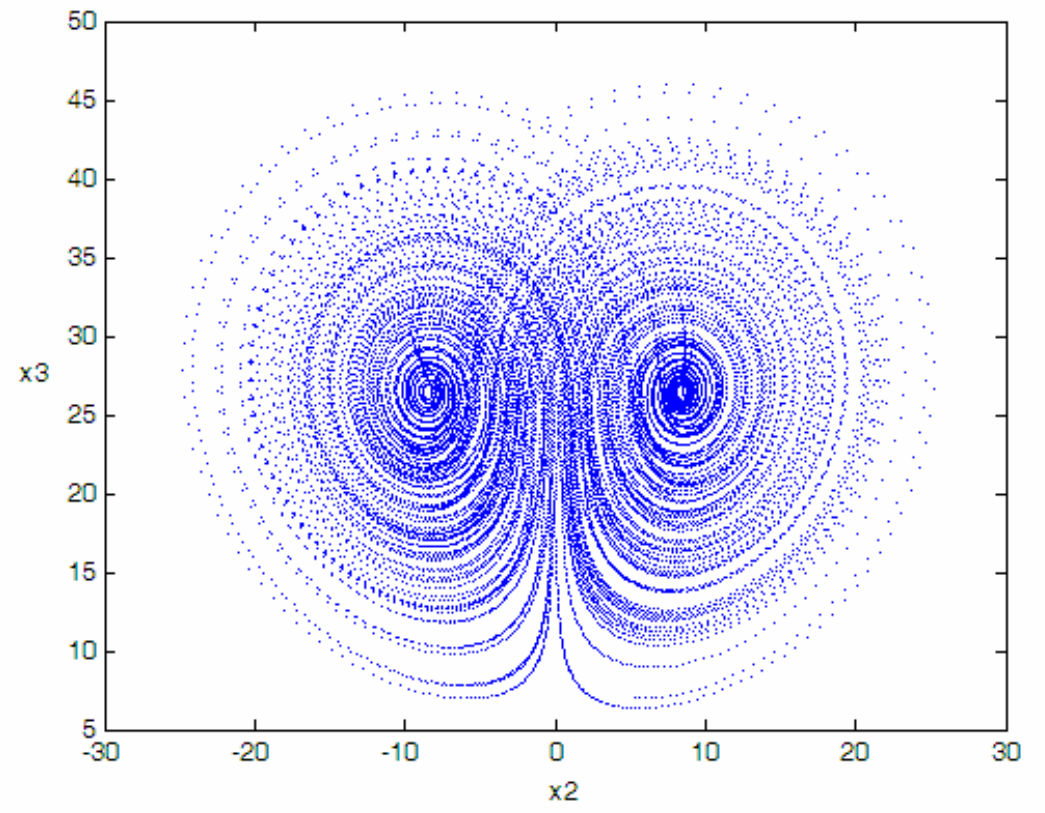

Fig. 14. The chaotic attractor of the unified chaotic system in the $x_2 - x_3$ plane, with parameters $\alpha = 0.1$, and initial condition $x(0) = (3,\ 5,\ 7)$.

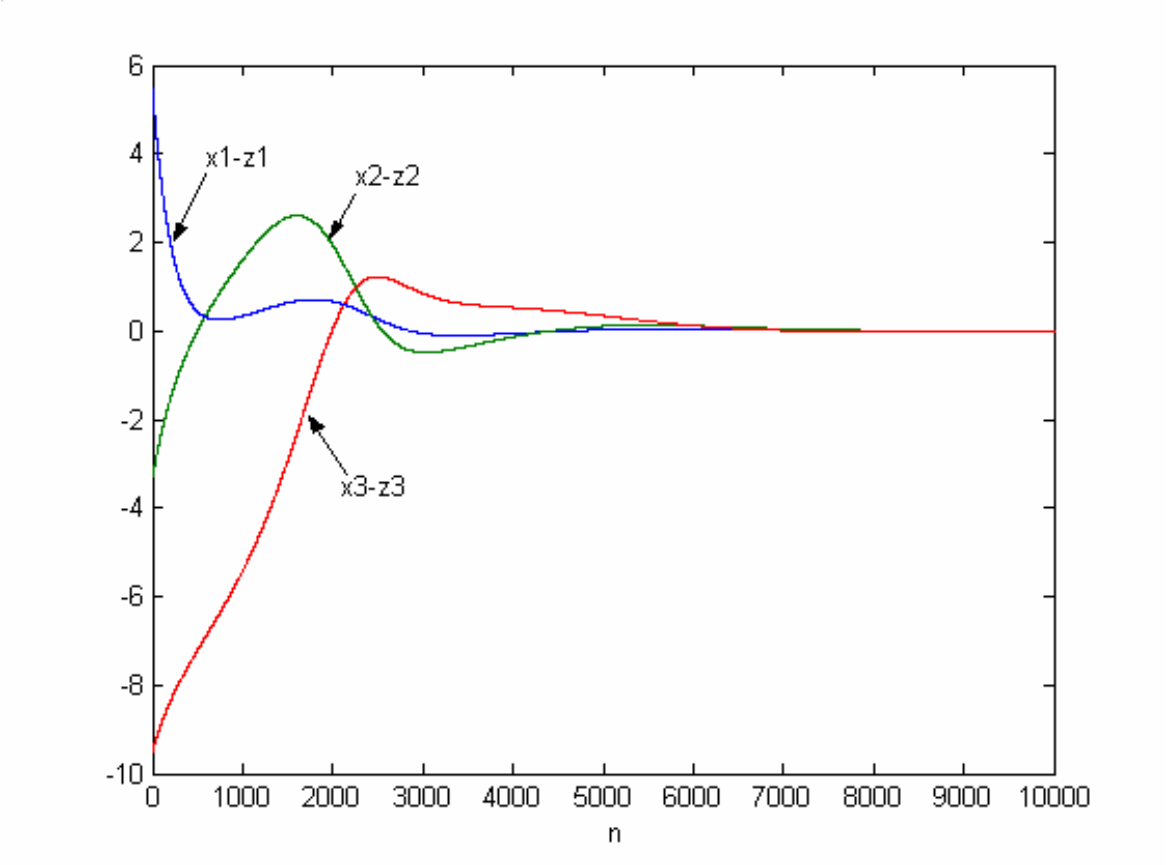

Fig. 15. Chaos synchronization of two unified chaotic systems coupled by $K = diag\{30, 15, 1.5\}$, with parameters $\alpha = 0.1$, and initial condition $x(0) = (3, 5, 7)$, $z(0) = (-2.5, 8.3, 16.5)$.

**Table**

Table 1. Comparison of the critical values.

| $\alpha$ | The critical value $k^*(6)$ | The critical value $k^{'}(6)$ |
|---|---|---|
| $\alpha = 0$ (Lorenz) | $39.2(k_{22}+1-29\alpha) + 352.5\left(k_{33}+\frac{8+\alpha}{3}\right)$ | $495.1(k_{22}+1-29\alpha) + 1701.6\left(k_{33}+\frac{8+\alpha}{3}\right)$ |
| $\alpha = 1$ (Chen) | $140.2(k_{22}+1-29\alpha) + 568.5\left(k_{33}+\frac{8+\alpha}{3}\right)$ | $484(k_{22}+1-29\alpha) + 1296\left(k_{33}+\frac{8+\alpha}{3}\right)$ |
| $\alpha = 0.8$ (Lü) | $115.4(k_{22}+1-29\alpha) + 525\left(k_{33}+\frac{8+\alpha}{3}\right)$ | $380.3(k_{22}+1-29\alpha) + 1190.3\left(k_{33}+\frac{8+\alpha}{3}\right)$ |